\documentclass[preprint,12pt]{elsarticle}
\usepackage{amsmath}
\usepackage{graphicx,psfrag,epsf}
\usepackage{enumerate}
\usepackage{natbib}
\usepackage{setspace}

\usepackage{amssymb}

\journal{arXiv}

\begin{document}

\begin{frontmatter}



\title{Robust designs to model uncertainty with high estimation and prediction efficiency}


\author{Chang-Yun Lin}

\address{Department of Applied Mathematics and Institute of Statistics,\\ National Chung Hsing University, Taiwan}

\begin{abstract}
Alphabetic optimality criteria, such as the $D$, $A$, and $I$ criteria, require specifying a model to select optimal designs. They are not model free and the optimal designs selected by them are not robust to model uncertainty. Recently, many extensions of the $D$ and $A$ criteria have been proposed for selecting robust designs with high estimation efficiency. However, approaches for finding robust designs with high prediction efficiency are rarely studied in the literature.  
In this paper, we propose the $P_\alpha$ criterion and develop its approximation version for two-level designs, called the ${\tilde P_\alpha}$ criterion. 
They are useful for selecting robust designs with high estimation, high prediction, or balanced estimation and prediction efficiency for projective submodels. 
Computational studies show that the ${\tilde P}_\alpha$ criterion is a good approximation of the $P_\alpha$ criterion and can reduce great computation time when we search designs over a wide range of models. 
The connection between the ${\tilde P_\alpha}$ criterion and the generalized minimum aberration (GMA) criterion is studied. 
Result shows that ${\tilde P_\alpha}$ plays a great role to link the alphabetic optimality criteria and the aberration-based criteria.
\end{abstract}


\end{frontmatter}



\section{Introduction}
It is an important task in the design of experiments to search for optimal designs that have good statistical properties. In the literature, there exist two main types of criteria that are commonly used for this purpose.  
%
The first type is the so-called model-free criterion, such as the minimum aberration (MA) criterion (Fries and Hunter, 1980), the minimum $G_2$ aberration (M$G_2$A) criterion (Tang and Dan, 1999), and the generalized minimum aberration (GMA) criterion (Xu and Wu, 2001). 
The MA criterion is a special case of the GMA criterion for regular designs and the M$G_2$A criterion is a special case of the GMA criterion for two-level orthogonal arrays. 
Designs selected by these criteria have minimum overall aliasing for lower-order effects.  %
The second type is the so-called alphabetic optimality criterion, such as the $D$, $A$, and $I$ criteria. 
The $D$ and $A$ criteria aim to minimize the variance of parameter estimation for the fitted model. Hence, designs selected by them have high estimation efficiency. The $I$-optimal criterion minimizes the average variance of predictions. Hence, 
designs selected by it have high prediction efficiency. 

The shortcoming of the alphabetic optimality criteria is that they are not model free. To select an optimal design using these criteria, experimenters need to specify a model that is assumed to be the final model and known in advance.  
This is impractical because the final fitted model is usually uncertain and almost never known. To tackle with model uncertainty, many extensions have been developed (L\"auter, 1974, Zhao et al., 2003, Heredia-Langner et al., 2004). 
As mentioned in Tsai and Gilmour (2010), these extensions only concentrate on small numbers of alternative models, which may not include the final model. 
Tsai et al. (2000) pointed out that only a few factors in an experiment are active and we end up fitting a model with a lower-dimensional projection of the original design. 
This is known as the effect sparsity principle (Box and Meyer, 1986). 
Therefore, a good design for experiments should be able to project onto good lower-dimensional designs for various sets of factors and provide good statistical properties for a range of possible models.
They introduced the concept of the maximal model of interest and assumed that the final fitted model would be a submodel of the maximal model. 
Based on this idea, Tsai et al. (2000) considered the average $A$-efficiency over all submodels of the maximal model, called the mean $A_s$-efficiency in their paper. Since it measures the overall mean of $A$-efficiency for all submodels, designs selected by it are robust to model uncertainty with high estimation efficiency. 
In practice, the number of the submodels is usually very large. It is sometimes difficult to calculate $A$-efficiency for all submodels. 
To overcome this difficulty, Tsai et al. (2000, 2007) developed the measures $Q$ and $Q_B$ to approximate the mean $A_s$-efficiency and the weighted mean $A_s$-efficiency, respectively. 
Since the two measures do not require the inversion of the information matrices of the submodels, they are computationally inexpensive and can play roles as the (weighted) mean $A_s$-efficiency to fast search for best robust designs. 

In this paper, we consider a practical situation in the data analysis that experimenters are also interested in prediction after the final model is fitted. 
Traditionally, the $I$ criterion is usually used to select designs with high prediction efficiency. 
As mentioned above, the $I$ criterion requires a pre-specified model and designs selected by it are not robust to model uncertainty. 
Unlike robust $D$-optimal and $A$-optimal designs, methods for finding robust $I$-optimal designs are rarely developed in the literature. 
Inspired by Tsai et al. (2000, 2007), we extend their methods to develop the $I_s$ criterion and its approximation version for two-level designs, called the ${\tilde I}_s$ criterion, to select designs that are robust to model uncertainty with high prediction efficiency. 
We further integrate these criteria to develop the $P_\alpha$ and ${\tilde P}_\alpha$ criteria, which are useful for selecting robust designs with high estimation, high prediction, or balanced estimation and prediction efficiency for projective submodels.     
The rest of this paper is organized as follows. Section~\ref{se:cr} introduces the criteria we proposed for selecting robust designs. Section~\ref{se:tr} discusses the connection between the ${\tilde P}_\alpha$ criterion and the GMA criterion. The applications of our method on regular, nonregular, saturated and supersaturated designs are given in Section~\ref{se:ap}. Section~\ref{se:co} is concluding remarks.

\section{Methods and Criteria}
\label{se:cr}

In this section, we introduce the method and the criteria in Tsai et al. (2000, 2007). 
By extending their method, we develop new criteria that are useful for selecting robust designs with high estimation and prediction efficiency. 

\subsection{The $A_s$, $I_s$, and $P_\alpha$ criteria}\label{se:p}
We first define the $A_s$ criterion, which is related to the (weighted) mean $A_s$-efficiency in Tsai et al. (2000, 2007).
Assume that a maximal model of interest includes $k$ main effects and $k^*$ of their two-factor interactions, where $k+k^*=v$.
Following the assumption in Tsai et al. (2000, 2007), the maximal model and its submodels must obey functional marginality (McCullagh and Nelder, 1989), called strong heredity by Chipman (1996), which means that every term in the model must be accompanied by all terms marginal to it. For instance, if interaction AB is in the model, then main effects A and B must be also included in the model. 
Consider the submodel of the maximal model, $M_s$, which includes $k_s$ ($\leq k$) main effects and $k_s^*$ ($\leq k^*$) two-factor interactions 
and express it by $E({\bf y})={\bf X}_s{\boldsymbol\beta}_s$, where $\bf y$ is an $N\times 1$ vector of responses, ${\boldsymbol\beta}_s=(\beta_0,\cdots,\beta_{v_s})'$ is a $(v_s+1)\times 1$ vector of parameters, and $\bf X_s$ is an $N\times (v_s+1)$ model matrix with the vector of 1's as its first column, followed by $k_s$ columns of main effects and $k_s^*$ columns of two-factor interactions, $k_s+k_s^*=v_s$. 
A model is inestimable when the number of parameters is greater than the run size. Hence, we call $M_s$ an eligible model if $(v_s+1)\leq N$ and an ineligible model if $(v_s+1)>N$. Note that the maximal model is not necessary to be an eligible model. 
Assume that there are total $n_0$ eligible submodels of the maximal model. Then the weighted average $A$-efficiency over the $n_0$ submodels can be measured by 
\begin{equation}\label{eq:As}
A_s=\sum_{s=1}^{n_0}tr[{\bf H}_s]p_s,
\end{equation}
where ${\bf H}_s$ is the $v_s\times v_s$ matrix obtained from $({\bf X}_s'{\bf X}_s)^{-1}$ by deleting its first row and first column, $tr[{\bf H}_s]$ is the trace of ${\bf H}_s$, and $p_s$ is the weight for the $s$th model with $\sum_{s=1}^{n_0}p_s=1$. 
Define the $A_s$ criterion, which selects a design with minimum $A_s$ value. 
Designs selected by the $A_s$ criterion are called the $A_s$-optimal designs, which are robust to model uncertainty with high estimation efficiency for submodels of the maximal model.    

With the same idea, we develop the $I_s$ criterion to evaluate designs with the weighted average $I$-efficiency over all eligible submodels.  
Let $f_s'(\cdot)$ be a model expansion function and $R_s=[-1,1]^{k_s}$ be the experimental region of the $k_s$ factors with respect to the submodel $M_s$. Denote $V[{\hat y}({\bf x}_s)]$ the variance of the predicted response for ${\bf x}_s\in R_s$. Then the weighted average $I$-efficiency over the $n_0$ eligible submodels can be measured by 
\begin{equation}\label{eq:v}
I_s=\sum_{s=1}^{n_0}\frac{\int_{R_s}V[{\hat y}({\bf x}_s)]d{\bf x}_s}{\int_{R_s}d{\bf x}_s}p_s=\sum_{s=1}^{n_0}\frac{\int_{R_s}f_s'({\bf x}_s)({\bf X}_s'{\bf X}_s)^{-1}f_s({\bf x}_s)d{\bf x}_s}{\int_{R_s}d{\bf x}_s}p_s.
\end{equation}
In Equation (\ref{eq:v}), the denominator, $\int_{R_s}d{\bf x}_s$, is the volume of the experimental region of the $k_s$ factors, which equals $2^{k_s}$, and the numerator can be simplified (see Jones and Goos, 2012) by 
\[
\begin{array}{l}
\int_{R_s}tr[f_s'({\bf x}_s)({\bf X}_s'{\bf X}_s)^{-1}f_s({\bf x}_s)]d{\bf x}_s\\
=\int_{R_s}tr[({\bf X}_s'{\bf X}_s)^{-1}f_s({\bf x}_s)f_s'({\bf x}_s)]d{\bf x}_s\\
=tr[({\bf X}_s'{\bf X}_s)^{-1}\int_{R_s}f_s({\bf x}_s)f_s'({\bf x}_s)d{\bf x}_s].
\end{array}
\]
Therefore, Equation (\ref{eq:v}) can be written as
\begin{equation}\label{eq:Is}
I_s=\sum_{s=1}^{n_0}tr[({\bf X}_s'{\bf X}_s)^{-1}{\bf G}_s]p_s,
\end{equation}
where 
\begin{equation}\label{eq:G}
{\bf G}_s=\frac{1}{2^{k_s}}\int_{R_s}f_s({\bf x}_s)f_s'({\bf x}_s)d{\bf x}_s=
\left[
\begin{array}{ccc}
1&{\bf 0}_{1\times k_s}&{\bf 0}_{1\times k_s^*}\\
{\bf 0}_{k_s\times 1}&\frac{1}{3}{\bf I}_{k_s}&{\bf 0}_{k_s\times k_s^*}\\
{\bf 0}_{k_s^*\times 1}&{\bf 0}_{k_s^*\times k_s}&\frac{1}{9}{\bf I}_{k_s^*}\\
\end{array}
\right].
\end{equation}
Define the $I_s$ criterion, which selects a design with minimum $I_s$ value. 
Designs selected by the $I_s$ criterion are called the $I_s$-optimal designs, which are robust to model uncertainty with high prediction efficiency for submodels of the maximal model.  

A design that has minimum $I_s$ value may not have minimum $A_s$ value, and vice versa. 
A trade-off between the two measures is given by 
\begin{equation}\label{eq:P}
P_\alpha=\alpha I_s+(1-\alpha)A_s,
\end{equation}
where $0\leq\alpha\leq 1$ controls the relative weight of $I_s$ and $A_s$ values. 
Define the $P_\alpha$ criterion, which selects a design with minimum $P_\alpha$ value. 
Designs selected by the $P_\alpha$ criterion are called the $P_\alpha$-optimal designs, which are robust to model uncertainty with balanced prediction and estimation efficiency for submodels of the maximal model.  
A $P_\alpha$-optimal design tends to have higher prediction efficiency when $\alpha$ is close to 1 and higher estimation efficiency when $\alpha$ is close to 0. Obviously, a $P_\alpha$-optimal design is an $A_s$-optimal design as $\alpha=0$ and an $I_s$-optimal design as $\alpha=1$.  
%

In practice, there may exist some eligible submodels that are inestimable. If this is the case, the following alternative measures which calculate the weighted harmonic mean are suggested: $A_s'=(\sum_{s=1}^{n_0}a_sp_s)^{-1}$, where $a_s=1/tr[{\bf H}_s]$ if $M_s$ is estimable and 0, otherwise, where ${\bf H}_s$ is defined in (\ref{eq:As}); $I_s'=(\sum_{s=1}^{n_0}e_sp_s)^{-1}$, where $e_s=1/tr[({\bf X_s}'{\bf X}_s) ^{-1}{\bf G}_s]$ if $M_s$ is estimable and 0, otherwise; and $P_\alpha'=\alpha I'_s+(1-\alpha)A_s'$.

\subsection{Approximate methods and the ${\tilde A}_s$, ${\tilde I}_s$, and ${\tilde P}_\alpha$ criteria}
\label{se:pa}
Calculations for the $A_s$, $I_s$, and $P_\alpha$ values require the inverses of the information matrices, i.e., $({\bf X}_s'{\bf X}_s)^{-1}$, for all eligible submodels. 
When the maximal model of interest is large, it is very time consuming to get all of the inverses matrices.  
%
Tsai et al. (2000, 2007) proposed the $Q$ and $Q_B$, which approximate the $A_s$ defined in Equation~(\ref{eq:As}).  
We extend their approach to derive the approximations for the $I_s$ and $P_\alpha$ defined in Equations~(\ref{eq:Is}) and (\ref{eq:P}). 

Let $\bf X$ be the model matrix of the maximal model. The first column of $\bf X$ is the vector of 1's, followed by $k$ columns of main effects and $k^*$ columns of two-factor interactions, where $v=k+k^*$. 
Let $a_{ij}$, $i,j=0,\cdots,v$, be the elements of the information matrix ${\bf X'X}$ and $c_{ii}$, $i=0,\cdots,v$, be the diagonal elements of $({\bf X'X})^{-1}$. 
According to the diagonal expansion for the determinant of matrices (Hohn 1973, p. 303) and applying the generalized Taylor's theorem by ignoring higher-order terms, $c_{ii}$ can be approximated (see Tsai et al., 2000) as 
\begin{equation}\label{eq:cii}
c_{ii}\approx\sum_{j=0}^vr_{ij},\mbox{ where }r_{ij}=\frac{1}{a_{ii}}\frac{a_{ij}^2}{a_{ii}a_{jj}}.
\end{equation}
The $I$-efficiency of the maximal design can be measured by 
\begin{equation}\label{eq:Iap}
tr[({\bf X'X})^{-1}{\bf G}]=c_{00}+\frac{1}{3}\sum_{i=1}^k c_{ii}+\frac{1}{9}\sum_{i=k+1}^vc_{ii},
\end{equation}
where $\bf G$ has the form in (\ref{eq:G}) as $k_s=k$ and $k_s^*=k^*$. 
Applying Equation~(\ref{eq:cii}), we obtain the approximation of Equation~(\ref{eq:Iap}) as 
\[
\sum_{j=0}^vr_{0j}+\frac{1}{3}\sum_{i=1}^k\sum_{j=0}^vr_{ij}+\frac{1}{9}\sum_{i=k+1}^v\sum_{j=0}^vr_{ij}.
\]
Following the notation in Tsai et al. (2000), for submodel $M_s$, define
\[
M_s(i,j)=\left\{
\begin{array}{ll}
1&\mbox{if effects }i\mbox{ and }j \mbox{ are both in }M_s,\\
0&\mbox{otherwise.} 
\end{array}
\right.
\]
Then the approximate $I$ value for submodel $M_s$ can be written as
\[
tr[({\bf X}_s'{\bf X}_s)^{-1}{\bf G}_s]\approx\sum_{j=0}^vr_{0j}M_s(0,j)+\frac{1}{3}\sum_{i=1}^k\sum_{j=0}^vr_{ij}M_s(i,j)+\frac{1}{9}\sum_{i=k+1}^v\sum_{j=0}^vr_{ij}M_s(i,j).
\]
Applying the above approximation to Equation (\ref{eq:Is}), we obtain the approximate $I_s$ value as  
\begin{equation}\label{eq:aIs}
{\tilde I}_s=\sum_{i=0}^v\sum_{j=0}^vg_{i}r_{ij}p_{ij}
\end{equation}
where $g_{i}=1\mbox{ for }i=0$, $g_{i}=1/3\mbox{ for }i=1\cdots,k$, $g_{i}=1/9\mbox{ for }i=k+1\cdots,v$, and $p_{ij}=\sum_{s=1}^{n_0}M_s(i,j)p_s$. 
The criterion which selects a design with minimum ${\tilde I}_s$ value is defined as the ${\tilde I}_s$ criterion. Designs selected by the ${\tilde I}_s$ criterion are called the ${\tilde I}_s$-optimal designs.

Similarly, the approximate $A_s$ can be obtained as
\begin{equation}\label{eq:aAs}
{\tilde A}_s=\sum_{i=1}^v\sum_{j=0}^vr_{ij}p_{ij}.
\end{equation}
The criterion which selects a design with minimum ${\tilde A}_s$ value is defined as the ${\tilde A}_s$ criterion. 
Designs selected by the ${\tilde A}_s$ criterion are called the ${\tilde A}_s$-optimal designs. 
The ${\tilde A}_s$ criterion is identical to the $Q$ criterion in Tsai et al. (2000) when $p_{ij}=\sum_{s=1}^{n_0}M_s(i,j)/n_0$ and to the $Q_B$ criterion in Tsai et al. (2007) when $p_{ij}$ are as specified in their paper. 

Since both ${\tilde I}_s$ and ${\tilde A}_s$ values are linear combinations of $r_{ij}p_{ij}$, the approximate $P_\alpha$ value can be obtained as
\begin{equation}\label{eq:aP}
{\tilde P}_\alpha=\alpha {\tilde I}_s+(1-\alpha){\tilde A}_s=\sum_{i=0}^v\sum_{j=0}^v\alpha_{i}r_{ij}p_{ij},
\end{equation}
where $\alpha_{i}=\alpha\mbox{ for }i=0$, $\alpha_{i}=1-2\alpha/3\mbox{ for }i=1,\cdots,k$, and $\alpha_{i}=1-8\alpha/9\mbox{ for }i=k+1\cdots,v$. 
The criterion which selects a design with minimum ${\tilde P}_\alpha$ value is defined as the ${\tilde P}_\alpha$ criterion. 
Designs selected by the ${\tilde P}_\alpha$ criterion are called the ${\tilde P}_\alpha$-optimal designs.


\subsection{Choice of $p_s$ and $p_{ij}$}
In this section, we discuss how to assign weights (or probabilities) to the submodels. 
A simple choice is equal weight, i.e., $p_s=1/n_0$ in Section~\ref{se:p}  and $p_{ij}=\sum_{s=1}^{n_0}M_s(i,j)/n_0$ in Section~\ref{se:pa}. The equal weight was used for the $Q$ criterion in Tsai et al. (2000). 
Now consider the situation that the experimenter has prior knowledge about the probability of each effect being in the final model. 
Under this situation, Tsai et al. (2007) extended the $Q$ criterion to the $Q_B$ criterion by assuming same prior probability for all main effects and same prior probability for all two-factor interactions. 
We relax this restriction and allow different prior probabilities assigned to different main effects and two-factor interactions. 
Let $\pi_i$ denote the experimenter's prior believe that the main effect of factor $i$ is in the final model, which is defined as
\begin{equation}\label{eq:pi1}
Pr(\delta_i=1)=\pi_i,
\end{equation}
where $\delta_i$ indicates whether or not the main effect of factor $i$ is in the final model. 
Under the functional marginality rule, we define $\pi_{ij}$ as the prior probability that the interaction of factors $i$ and $j$ is in the final model given that the main effects of both factors $i$ and $j$ are in the model, so that
\begin{equation}\label{eq:pi2}
Pr(\delta_{ij}=1|\delta_i,\delta_j)=\left\{\begin{array}{ll}
\pi_{ij}&\mbox{ if }\delta_{i}=\delta_{j}=1,\\
0&\mbox{ otherwise.}\\
\end{array}
\right.
\end{equation}
Then, the probability of model $M_s$ being the final model is  
\[
Pr(M_s)=\prod_{i=1}^k\pi_i^{\delta_i}(1-\pi_i)^{k-\delta_i}\prod_{i=1}^{k-1}\prod_{j=i+1}^k\pi_{ij}^{\delta_{ij}}(1-\pi_{ij})^{1-\delta_{ij}}.
\]
If the number of parameters in $M_s$ is greater than the run size, i.e., $(v_s+1)>N$, then $M_s$ is ineligible. Hence, we adjust the probability by
\begin{equation}\label{eq:adj}
\tilde{Pr}(M_s)=\left\{
\begin{array}{ll}
Pr(M_s)/\gamma&\mbox{ if }(v_s+1) \leq N,\\
0&\mbox{ if }(v_s+1)>N,\\  
\end{array}
\right. 
\end{equation}
where $\gamma$ is the sum of the probabilities for all eligible submodels. 
According to Equation~(\ref{eq:adj}), experimenters can calculate the $A_s$, $I_s$, $P_\alpha$ values by setting $p_s=Pr(M_s)/\gamma$ and calculate the ${\tilde A}_s$, ${\tilde I}_s$, ${\tilde P}_\alpha$ values by setting $p_{ij}=\sum_{s=1}^{n_0}M_s(i,j)Pr(M_s)/\gamma$. 
Note that, Equation (\ref{eq:adj}) is different from the adjustment given in Tsai et al. (2007). In their paper, the probabilities are adjusted only for the eligible submodels with $(v_s+1) = N$ but not for those with $(v_s+1) < N$. If there are many ineligible submodels (it happens when $k$ is close to $N$), then the adjusted probabilities of the $M_s$ with $(v_s+1) = N$ will become unreasonably larger than the probabilities of the $M_s$ with $(v_s+1)<N$. 
Our method adjusts the probability for all eligible submodels by multiplying $1/\gamma$, which maintains their proportion and forces them sum to one. 


\section{Connection between the ${\tilde P}_\alpha$ and GMA criteria}\label{se:tr}
In this section, we discuss the connection between the ${\tilde P}_\alpha$ criterion and the generalized word counts in the GMA criterion. 
Note that the ${\tilde I}_s$ and ${\tilde A}_s$ criteria are special cases of the ${\tilde P}_\alpha$ criterion when $\alpha=1$ and 0, respectively. 
Let $D=(d_{ij})$ represent an $N\times k$ fractional factorial design, where $d_{ij}$ is the level (coded by 1 and -1) of the $j$th factor in the $i$th run, $i=1,\cdots,N$, $j=1,\cdots,k$. 
For $1\leq l\leq k$ and any $l$-subset $w=\{{j_1},\cdots,{j_l}\}$ of $\{1,\cdots,k\}$, define 
$J_l(w)=\left|\sum_{i=1}^Nd_{ij_1}\cdots d_{ij_l}\right|$. 
It is the $J$-characteristic defined in Tang and Dang (1999). 
The generalized word count in Tsai and Gilmour (2010) for the words referring to the effects of $l$ factors can be defined by 
\[
b_l=\sum_{\|w\|=l}[J_l(w)/N]^2,
\]
where $\|w\|$ denotes the number of elements in $w$. It measures the overall aliasing between the intercept and main effects as $l=1$ and the overall aliasing between the intercept and interactions of $l$ factors as $l\geq 2$. 
For $l=0$, define $b_0=N$. 
The generalized wordlength pattern (GWLP) of $D$ is defined as $W(D)=(b_1,b_2,\cdots,b_k)$. The GMA criterion selects a design by sequentially minimizing $b_1,b_2,\cdots,b_k$.   

\subsection{First-order model}\label{se:1st}
When the maximal model of interest is a first-order model including the main effects of $k$ factors, i.e., $k^*=0$ and $v=k$,  
the ${\tilde P}_\alpha$ criterion is equivalent to selecting a design that minimizes
\begin{equation}\label{eq:w1}
\alpha\sum_{j=1}^kp_{0j}\frac{a_{0j}^2}{N^2}+(1-\frac{2}{3}\alpha)\sum_{i=1}^kp_{i0}\frac{a_{i0}^2}{N^2}+(1-\frac{2}{3}\alpha)\sum_{i=1}^k\sum_{\substack{j=1\\i\neq j}}^kp_{ij}\frac{a_{ij}^2}{N^2}.
\end{equation}
Following the notations and discussions in Tsai and Gilmour (2010), assume that $p_{ij}=\xi_2$, $\forall i,j,i\neq j$, and that $p_{i0}=\xi_1$ for $i=1,\cdots,k$, which also implies $p_{0j}=\xi_1$ for $j=1,\cdots,k$. Since $b_1=\sum_{i=1}^ka_{i0}^2/N^2=\sum_{j=1}^ka_{0j}^2/N^2$ and $2b_2=\sum_{i=1}^k\sum_{\substack{j=1\\i\neq j}}^ka_{ij}^2/N^2$, Equation (\ref{eq:w1}) can be written as 
\begin{equation}\label{eq:1st}
(1+\frac{\alpha}{3})\xi_1b_1+2(1-\frac{2}{3}\alpha)\xi_2b_2.
\end{equation}
When $(1+\alpha/3)\xi_1$ is much greater than $2(1-2\alpha/3)\xi_2$, the ${\tilde P}_\alpha$ criterion is equivalent to selecting a design by sequentially minimizing $b_1$ and $b_2$.  

\subsection{Second-order model}\label{se:2nd}
When the maximal model of interest is a second-order model which includes $k$ main effects and $k^*={k\choose 2}$ two-factor interactions, i.e., $v=k+k^*$. 
Assume that $\pi_i=\pi_1$ for $i=1,\cdots,k$ in (\ref{eq:pi1}) and $\pi_{ij}=\pi_2$ for $i,j=1,\cdots,k$ in (\ref{eq:pi2}).  
Following the notation in Tsai and Gilmour (2010), 
let $\xi_{ij}$ denote the sum of prior probabilities of the models, which include $i$ main effects and $j$ two-factor interactions, being the final model. 
Then the ${\tilde P}_\alpha$ criterion is equivalent to selecting a design that minimizes
\begin{equation}\label{eq:2nd}
\begin{array}{l}
[(1+\frac{\alpha}{3})\xi_{10}+2(1-\frac{7\alpha}{9})(k-1)\xi_{21}]b_1+[2(1-\frac{2\alpha}{3})\xi_{20}+(1+\frac{\alpha}{9})\xi_{21}+2(1-\frac{8\alpha}{9})(k-2)\xi_{32}]b_2\\
+6(1-\frac{7\alpha}{9})\xi_{31}b_3+6(1-\frac{8\alpha}{9})\xi_{42}b_4.
\end{array}
\end{equation}
Details of this result are given in Appendix A. 
It reduces to Equation (5) in Tsai and Gilmour (2010) when $\alpha=0$. 
For orthogonal arrays of strength 2, we have $b_1=b_2=0$ and, hence, Equation (\ref{eq:2nd}) reduces to 
\[
6(1-\frac{7}{9}\alpha)\xi_{31}b_3+6(1-\frac{8}{9}\alpha)\xi_{42}b_4.
\]
Since ${\tilde P}_\alpha$ is an approximation of $P_\alpha$ and has close relationship with the generalized word count, 
the ${\tilde P}_\alpha$ criterion establishes a great link between the alphabetic optimality criteria and the aberration-based criteria.  

\section{Applications}
\label{se:ap}
In this section, we provide several examples to illustrate the applications of our method to find robust regular, nonregular, saturated and supersaturated designs that are estimation and prediction efficient for projective submodels.

\subsection{Regular designs}
In regular designs, two effects (or the intercept and effects) are either uncorrelated or fully aliasing, which leads to $a_{ij}$ being either 0 or $\pm N$. 
When $M_s$ is estimable, $a_{ij}=0$ for all $i,j,i\neq j$ referring to the intercept and effects in $M_s$ and, hence, the approximation in Equation (\ref{eq:cii}) is the exact value of $c_{ii}$.    
Therefore, for a regular design, if the maximal model is estimable, then ${\tilde P}_\alpha=P_\alpha$.

{\bf Example 4.1.1.} For fractional factorial design $2^{k-p}_{III}$, where the subscript denotes the resolution, we have $b_1=b_2=0$. Hence, Equation (\ref{eq:1st}) equals 0, which is minimum for any $\alpha$ between 0 and 1.  
According to Section~\ref{se:1st}, if the maximal model of interest is a first-order model, then design $2^{k-p}_{III}$ is a ${\tilde P}_\alpha$-optimal design. 
Since any first-order model of resolution-III designs is estimable, we have ${\tilde P}_\alpha=P_\alpha$ and, hence, $2^{k-p}_{III}$ is also a $P_\alpha$-optimal design.   
Therefore, regular design $2^{k-p}_{III}$ is robust to first-order models with highest estimation and prediction efficiency. 
%

{\bf Example 4.1.2.} For fractional factorial design $2^{k-p}_{V}$, we have $b_l=0$ for $l=1,\cdots,4$. Hence, Equation (\ref{eq:2nd}) equals 0. According to Section~\ref{se:2nd}, if the maximal model of interest is a second-order model, then design $2^{k-p}_{V}$ is a ${\tilde P}_\alpha$-optimal design. Since any eligible second-order model of resolution-V designs is estimable, we have ${\tilde P}_\alpha=P_\alpha$ and, hence, $2^{k-p}_{V}$ is also a $P_\alpha$-optimal design.
Therefore, when the maximal model of interest is an estimable second-order model, regular design $2^{k-p}_{V}$ is a robust design with highest estimation and prediction efficiency for projective submodels. 

\begin{table}
\begin{center}
\caption{Four regular $16\times 5$ designs in Example~4.1.3.}\label{tb:ex4.1.3}
\begin{spacing}{.7}
\begin{tabular}{cccc}									
\begin{tabular}{rrrrr}									
\multicolumn{5}{c}{$A_1$}\\									
\hline									
1	&	2	&	3	&	4	&	5	\\
\hline									
1	&	1	&	1	&	1	&	1	\\
1	&	1	&	1	&	1	&	1	\\
1	&	1	&	1	&	-1	&	-1	\\
1	&	1	&	1	&	-1	&	-1	\\
1	&	-1	&	-1	&	1	&	1	\\
1	&	-1	&	-1	&	1	&	1	\\
1	&	-1	&	-1	&	-1	&	-1	\\
1	&	-1	&	-1	&	-1	&	-1	\\
-1	&	1	&	-1	&	1	&	-1	\\
-1	&	1	&	-1	&	1	&	-1	\\
-1	&	1	&	-1	&	-1	&	1	\\
-1	&	1	&	-1	&	-1	&	1	\\
-1	&	-1	&	1	&	1	&	-1	\\
-1	&	-1	&	1	&	1	&	-1	\\
-1	&	-1	&	1	&	-1	&	1	\\
-1	&	-1	&	1	&	-1	&	1	\\
\hline									
\end{tabular}									
&									
\begin{tabular}{rrrrr}									
\multicolumn{5}{c}{$A_2$}\\									
\hline									
1	&	2	&	3	&	4	&	5	\\
\hline									
1	&	1	&	1	&	1	&	1	\\
1	&	1	&	1	&	1	&	-1	\\
1	&	1	&	1	&	-1	&	1	\\
1	&	1	&	1	&	-1	&	-1	\\
1	&	-1	&	-1	&	1	&	1	\\
1	&	-1	&	-1	&	1	&	-1	\\
1	&	-1	&	-1	&	-1	&	1	\\
1	&	-1	&	-1	&	-1	&	-1	\\
-1	&	1	&	-1	&	1	&	1	\\
-1	&	1	&	-1	&	1	&	-1	\\
-1	&	1	&	-1	&	-1	&	1	\\
-1	&	1	&	-1	&	-1	&	-1	\\
-1	&	-1	&	1	&	1	&	1	\\
-1	&	-1	&	1	&	1	&	-1	\\
-1	&	-1	&	1	&	-1	&	1	\\
-1	&	-1	&	1	&	-1	&	-1	\\
\hline									
\end{tabular}									
&									
\begin{tabular}{rrrrr}									
\multicolumn{5}{c}{$A_3$}\\									
\hline									
1	&	2	&	3	&	4	&	5	\\
\hline									
1	&	1	&	1	&	1	&	1	\\
1	&	1	&	1	&	1	&	-1	\\
1	&	1	&	-1	&	-1	&	1	\\
1	&	1	&	-1	&	-1	&	-1	\\
1	&	-1	&	1	&	-1	&	1	\\
1	&	-1	&	1	&	-1	&	-1	\\
1	&	-1	&	-1	&	1	&	1	\\
1	&	-1	&	-1	&	1	&	-1	\\
-1	&	1	&	1	&	-1	&	1	\\
-1	&	1	&	1	&	-1	&	-1	\\
-1	&	1	&	-1	&	1	&	1	\\
-1	&	1	&	-1	&	1	&	-1	\\
-1	&	-1	&	1	&	1	&	1	\\
-1	&	-1	&	1	&	1	&	-1	\\
-1	&	-1	&	-1	&	-1	&	1	\\
-1	&	-1	&	-1	&	-1	&	-1	\\
\hline									
\end{tabular}									
&									
\begin{tabular}{rrrrr}									
\multicolumn{5}{c}{$A_4$}\\									
\hline									
1	&	2	&	3	&	4	&	5	\\
\hline									
1	&	1	&	1	&	1	&	1	\\
1	&	1	&	1	&	-1	&	-1	\\
1	&	1	&	-1	&	1	&	-1	\\
1	&	1	&	-1	&	-1	&	1	\\
1	&	-1	&	1	&	1	&	-1	\\
1	&	-1	&	1	&	-1	&	1	\\
1	&	-1	&	-1	&	1	&	1	\\
1	&	-1	&	-1	&	-1	&	-1	\\
-1	&	1	&	1	&	1	&	-1	\\
-1	&	1	&	1	&	-1	&	1	\\
-1	&	1	&	-1	&	1	&	1	\\
-1	&	1	&	-1	&	-1	&	-1	\\
-1	&	-1	&	1	&	1	&	1	\\
-1	&	-1	&	1	&	-1	&	-1	\\
-1	&	-1	&	-1	&	1	&	-1	\\
-1	&	-1	&	-1	&	-1	&	1	\\
\hline									
\end{tabular}									
\end{tabular}									
\end{spacing}
\end{center}
\end{table}

{\bf Example 4.1.3.} 
Table~\ref{tb:ex4.1.3} lists four nonisomorphic regular designs $A_1,\cdots,A_4$, which are obtained from the projections of  the $16\times 15$ design in Table~5 in Tsai and Gilmour (2010). 
We first use the GMA criterion to select the best design among them. 
The generalized wordlength patterns of the four designs are $W(A_1)=(0,0,2,1,0)$, $W(A_2)=(0,0,1,0,0)$, $W(A_3)=(0,0,0,1,0)$, and $W(A_4)=(0,0,0,0,1)$.  
According to the GMA criterion, we obtain $A_4\gg A_3\gg A_2\gg A_1$, where ``$X \gg Y$" indicates that $X$ is better than $Y$. 
Now, we apply the ${\tilde P}_\alpha$ criterion by considering equal weight to all submodels and setting $\alpha=0.5$. Assume that the maximal model of interest is the second-order model including main effects of five factors and all of their two-factor interactions. 
The ${\tilde P}_\alpha$ values of the four designs are ${\tilde P}_{.5}(A_1)=0.5945$, ${\tilde P}_{.5}(A_2)=0.4637$, ${\tilde P}_{.5}(A_3)=0.4111$, and ${\tilde P}_{.5}(A_4)=0.3721$. 
According to the ${\tilde P}_\alpha$ criterion, we obtain $A_4\gg A_3\gg A_2\gg A_1$. 
In this example, the two criteria have the same result and select $A_4$ as the best design. 
Since $A_4$ has resolution $V$, it has minimum aberration and is robust to model uncertainty with highest estimation and prediction efficiency.  

\subsection{Nonregular designs}
In nonregular designs, one effect/interaction may partially alias with another effect/interaction or the intercept. Hence, $a_{ij}$, $i\neq j$, in nonregular deigns are between $-N$ and $N$. If
effects in submodels are partial aliasing, then ${\tilde P}_\alpha$ does not equal $P_\alpha$. 
In the following example, we calculate the approximate values, ${\tilde P}_\alpha$, and exact values, $P_\alpha$, of twelve nonregular designs. 
We compare the computation time between the two measures and examine whether ranking designs by the ${\tilde P}_\alpha$ and $P_\alpha$ criteria is consistent.

\begin{table}
\begin{center}
\caption{Twelve $14\times 5$ nonregular designs in Example 4.2.1.}\label{tb:ex4.2.1}
\begin{spacing}{.7}
\begin{tabular}{cccc}									
\begin{tabular}{rrrrr}									
\multicolumn{5}{c}{$B_1$}\\									
\hline									
1	&	2	&	3	&	4	&	5	\\
\hline									
-1	&	1	&	1	&	-1	&	1	\\
1	&	-1	&	-1	&	-1	&	-1	\\
1	&	-1	&	-1	&	-1	&	1	\\
1	&	-1	&	-1	&	1	&	1	\\
-1	&	1	&	-1	&	-1	&	1	\\
-1	&	-1	&	1	&	-1	&	-1	\\
-1	&	1	&	1	&	1	&	-1	\\
1	&	1	&	1	&	1	&	1	\\
-1	&	-1	&	1	&	-1	&	1	\\
-1	&	-1	&	-1	&	1	&	-1	\\
-1	&	1	&	-1	&	1	&	1	\\
1	&	-1	&	1	&	1	&	-1	\\
1	&	1	&	1	&	-1	&	-1	\\
1	&	1	&	-1	&	1	&	-1	\\
\hline									
\end{tabular}									
&									
\begin{tabular}{rrrrr}									
\multicolumn{5}{c}{$B_2$}\\									
\hline									
1	&	2	&	3	&	4	&	5	\\
\hline									
-1	&	1	&	1	&	-1	&	1	\\
1	&	-1	&	-1	&	1	&	-1	\\
1	&	-1	&	-1	&	1	&	1	\\
1	&	-1	&	-1	&	-1	&	1	\\
-1	&	1	&	-1	&	-1	&	1	\\
-1	&	-1	&	1	&	1	&	-1	\\
-1	&	1	&	1	&	1	&	-1	\\
1	&	1	&	1	&	1	&	1	\\
-1	&	-1	&	1	&	-1	&	1	\\
-1	&	-1	&	-1	&	-1	&	-1	\\
-1	&	1	&	-1	&	1	&	1	\\
1	&	-1	&	1	&	1	&	-1	\\
1	&	1	&	1	&	-1	&	-1	\\
1	&	1	&	-1	&	-1	&	-1	\\
\hline									
\end{tabular}									
&									
\begin{tabular}{rrrrr}									
\multicolumn{5}{c}{$B_3$}\\									
\hline									
1	&	2	&	3	&	4	&	5	\\
\hline									
-1	&	1	&	1	&	-1	&	1	\\
1	&	-1	&	-1	&	-1	&	-1	\\
1	&	-1	&	-1	&	1	&	1	\\
1	&	-1	&	-1	&	-1	&	1	\\
-1	&	1	&	-1	&	-1	&	1	\\
-1	&	-1	&	1	&	1	&	-1	\\
-1	&	1	&	1	&	-1	&	-1	\\
1	&	1	&	1	&	1	&	1	\\
-1	&	-1	&	1	&	1	&	1	\\
-1	&	-1	&	-1	&	-1	&	-1	\\
-1	&	1	&	-1	&	1	&	1	\\
1	&	-1	&	1	&	-1	&	-1	\\
1	&	1	&	1	&	1	&	-1	\\
1	&	1	&	-1	&	1	&	-1	\\
\hline									
\end{tabular}									
&									
\begin{tabular}{rrrrr}									
\multicolumn{5}{c}{$B_4$}\\									
\hline									
1	&	2	&	3	&	4	&	5	\\
\hline									
-1	&	1	&	1	&	-1	&	-1	\\
1	&	-1	&	-1	&	-1	&	1	\\
1	&	-1	&	-1	&	-1	&	-1	\\
1	&	-1	&	-1	&	1	&	-1	\\
-1	&	1	&	-1	&	-1	&	1	\\
-1	&	-1	&	1	&	-1	&	-1	\\
-1	&	1	&	1	&	1	&	-1	\\
1	&	1	&	1	&	1	&	1	\\
-1	&	-1	&	1	&	-1	&	1	\\
-1	&	-1	&	-1	&	1	&	1	\\
-1	&	1	&	-1	&	1	&	-1	\\
1	&	-1	&	1	&	1	&	1	\\
1	&	1	&	1	&	-1	&	1	\\
1	&	1	&	-1	&	1	&	-1	\\
\hline									
\end{tabular}\\									
&&&\\									
\begin{tabular}{rrrrr}									
\multicolumn{5}{c}{$B_5$}\\									
\hline									
1	&	2	&	3	&	4	&	5	\\
\hline									
-1	&	1	&	1	&	1	&	-1	\\
1	&	-1	&	-1	&	-1	&	-1	\\
1	&	-1	&	-1	&	-1	&	-1	\\
1	&	-1	&	-1	&	-1	&	1	\\
-1	&	1	&	-1	&	1	&	-1	\\
-1	&	-1	&	1	&	1	&	-1	\\
-1	&	1	&	1	&	-1	&	1	\\
1	&	1	&	1	&	1	&	1	\\
-1	&	-1	&	1	&	-1	&	-1	\\
-1	&	-1	&	-1	&	1	&	1	\\
-1	&	1	&	-1	&	-1	&	1	\\
1	&	-1	&	1	&	1	&	1	\\
1	&	1	&	1	&	-1	&	-1	\\
1	&	1	&	-1	&	1	&	1	\\
\hline									
\end{tabular}									
&									
\begin{tabular}{rrrrr}									
\multicolumn{5}{c}{$B_6$}\\									
\hline									
1	&	2	&	3	&	4	&	5	\\
\hline									
-1	&	1	&	1	&	-1	&	-1	\\
1	&	-1	&	-1	&	1	&	1	\\
1	&	-1	&	-1	&	-1	&	1	\\
1	&	-1	&	-1	&	-1	&	-1	\\
-1	&	1	&	-1	&	1	&	-1	\\
-1	&	-1	&	1	&	-1	&	1	\\
-1	&	1	&	1	&	-1	&	1	\\
1	&	1	&	1	&	1	&	1	\\
-1	&	-1	&	1	&	1	&	-1	\\
-1	&	-1	&	-1	&	1	&	-1	\\
-1	&	1	&	-1	&	-1	&	1	\\
1	&	-1	&	1	&	1	&	1	\\
1	&	1	&	1	&	1	&	-1	\\
1	&	1	&	-1	&	-1	&	-1	\\
\hline									
\end{tabular}									
&									
\begin{tabular}{rrrrr}									
\multicolumn{5}{c}{$B_7$}\\									
\hline									
1	&	2	&	3	&	4	&	5	\\
\hline									
-1	&	1	&	1	&	1	&	-1	\\
1	&	-1	&	-1	&	-1	&	1	\\
1	&	-1	&	-1	&	-1	&	1	\\
1	&	-1	&	-1	&	-1	&	-1	\\
-1	&	1	&	-1	&	1	&	-1	\\
-1	&	-1	&	1	&	1	&	1	\\
-1	&	1	&	1	&	-1	&	1	\\
1	&	1	&	1	&	1	&	1	\\
-1	&	-1	&	1	&	-1	&	-1	\\
-1	&	-1	&	-1	&	1	&	-1	\\
-1	&	1	&	-1	&	-1	&	1	\\
1	&	-1	&	1	&	1	&	1	\\
1	&	1	&	1	&	-1	&	-1	\\
1	&	1	&	-1	&	1	&	-1	\\
\hline									
\end{tabular}									
&									
\begin{tabular}{rrrrr}									
\multicolumn{5}{c}{$B_8$}\\									
\hline									
1	&	2	&	3	&	4	&	5	\\
\hline									
-1	&	1	&	1	&	-1	&	-1	\\
1	&	-1	&	-1	&	-1	&	-1	\\
1	&	-1	&	-1	&	-1	&	1	\\
1	&	-1	&	-1	&	1	&	-1	\\
-1	&	1	&	-1	&	-1	&	-1	\\
-1	&	-1	&	1	&	-1	&	1	\\
-1	&	1	&	1	&	1	&	-1	\\
1	&	1	&	1	&	1	&	1	\\
-1	&	-1	&	1	&	-1	&	1	\\
-1	&	-1	&	-1	&	1	&	-1	\\
-1	&	1	&	-1	&	1	&	1	\\
1	&	-1	&	1	&	1	&	-1	\\
1	&	1	&	1	&	-1	&	1	\\
1	&	1	&	-1	&	1	&	1	\\
\hline									
\end{tabular}\\									
&&&\\									
\begin{tabular}{rrrrr}									
\multicolumn{5}{c}{$B_9$}\\									
\hline									
1	&	2	&	3	&	4	&	5	\\
\hline									
-1	&	1	&	1	&	-1	&	1	\\
1	&	-1	&	-1	&	1	&	1	\\
1	&	-1	&	-1	&	-1	&	1	\\
1	&	-1	&	-1	&	-1	&	-1	\\
-1	&	1	&	-1	&	1	&	-1	\\
-1	&	-1	&	1	&	-1	&	-1	\\
-1	&	1	&	1	&	-1	&	1	\\
1	&	1	&	1	&	1	&	1	\\
-1	&	-1	&	1	&	1	&	1	\\
-1	&	-1	&	-1	&	1	&	1	\\
-1	&	1	&	-1	&	-1	&	-1	\\
1	&	-1	&	1	&	1	&	-1	\\
1	&	1	&	1	&	1	&	-1	\\
1	&	1	&	-1	&	-1	&	-1	\\
\hline									
\end{tabular}									
&									
\begin{tabular}{rrrrr}									
\multicolumn{5}{c}{$B_{10}$}\\									
\hline									
1	&	2	&	3	&	4	&	5	\\
\hline									
-1	&	1	&	1	&	-1	&	-1	\\
1	&	-1	&	-1	&	-1	&	1	\\
1	&	-1	&	-1	&	1	&	-1	\\
1	&	-1	&	-1	&	-1	&	1	\\
-1	&	1	&	-1	&	-1	&	1	\\
-1	&	-1	&	1	&	1	&	1	\\
-1	&	1	&	1	&	-1	&	-1	\\
1	&	1	&	1	&	1	&	1	\\
-1	&	-1	&	1	&	1	&	1	\\
-1	&	-1	&	-1	&	-1	&	-1	\\
-1	&	1	&	-1	&	1	&	-1	\\
1	&	-1	&	1	&	-1	&	-1	\\
1	&	1	&	1	&	1	&	-1	\\
1	&	1	&	-1	&	1	&	1	\\
\hline									
\end{tabular}									
&									
\begin{tabular}{rrrrr}									
\multicolumn{5}{c}{$B_{11}$}\\									
\hline									
1	&	2	&	3	&	4	&	5	\\
\hline									
-1	&	1	&	-1	&	-1	&	-1	\\
1	&	-1	&	-1	&	-1	&	1	\\
1	&	-1	&	-1	&	1	&	1	\\
1	&	-1	&	1	&	-1	&	-1	\\
-1	&	1	&	-1	&	-1	&	-1	\\
-1	&	-1	&	-1	&	1	&	1	\\
-1	&	1	&	1	&	-1	&	1	\\
1	&	1	&	1	&	1	&	1	\\
-1	&	-1	&	-1	&	1	&	-1	\\
-1	&	-1	&	1	&	-1	&	-1	\\
-1	&	1	&	1	&	1	&	1	\\
1	&	-1	&	1	&	-1	&	1	\\
1	&	1	&	-1	&	1	&	-1	\\
1	&	1	&	1	&	1	&	-1	\\
\hline									
\end{tabular}									
&									
\begin{tabular}{rrrrr}									
\multicolumn{5}{c}{$B_{12}$}\\									
\hline									
1	&	2	&	3	&	4	&	5	\\
\hline									
-1	&	1	&	1	&	-1	&	-1	\\
1	&	-1	&	-1	&	1	&	-1	\\
1	&	-1	&	-1	&	-1	&	1	\\
1	&	-1	&	-1	&	-1	&	-1	\\
-1	&	1	&	-1	&	1	&	-1	\\
-1	&	-1	&	1	&	-1	&	1	\\
-1	&	1	&	1	&	-1	&	-1	\\
1	&	1	&	1	&	1	&	1	\\
-1	&	-1	&	1	&	1	&	1	\\
-1	&	-1	&	-1	&	1	&	-1	\\
-1	&	1	&	-1	&	-1	&	1	\\
1	&	-1	&	1	&	1	&	-1	\\
1	&	1	&	1	&	1	&	1	\\
1	&	1	&	-1	&	-1	&	1	\\
\hline									
\end{tabular}\\									
\end{tabular}																	
\end{spacing}
\end{center}
\end{table}

\begin{table}
\begin{center}
\caption{Comparison of the $P_\alpha$ and ${\tilde P}_\alpha$ criteria for twelve $14\times 5$ nonregular designs in Table~\ref{tb:ex4.2.1}.}\label{tb:non}
\begin{tabular}{c|cccccccc}																																																													
\hline																																																													
Max. mod.	&		\multicolumn{2}{c}{$k=2$}										&		\multicolumn{2}{c}{$k=3$}														&		\multicolumn{2}{c}{$k=4$}														&		\multicolumn{2}{c}{$k=5$}														\\
Criterion	&		${{P}}_{\alpha}$				&		${{\tilde P}}_{\alpha}$				&		${{P}}_{\alpha}$						&		${{\tilde P}}_{\alpha}$						&		${{P}}_{\alpha}$						&		${{\tilde P}}_{\alpha}$						&		${{P}}_{\alpha}$						&		${{\tilde P}}_{\alpha}$						\\
\hline																																																													
$B_1$	&	$	.1019	^	1	$	&	$	.1018	^	1	$	&	$	.1799	^	{	1	}	$	&	$	.1789	^	{	1	}	$	&	$	.2574	^	{	1	}	$	&	$	.3109	^	{	1	}	$	&	$	.5132	^	{	1	}	$	&	$	.5087	^	{	1	}	$	\\
$B_2$	&	$	.1019	^	1	$	&	$	.1018	^	1	$	&	$	.1809	^	{	2	}	$	&	$	.1798	^	{	2	}	$	&	$	.2609	^	{	2	}	$	&	$	.3174	^	{	2	}	$	&	$	.5559	^	{	2	}	$	&	$	.5328	^	{	2	}	$	\\
$B_3$	&	$	.1019	^	1	$	&	$	.1018	^	1	$	&	$	.1850	^	{	3	}	$	&	$	.1808	^	{	3	}	$	&	$	.2674	^	{	3	}	$	&	$	.3218	^	{	3	}	$	&	$	.5845	^	{	3	}	$	&	$	.5426	^	{	3	}	$	\\
$B_4$	&	$	.1019	^	1	$	&	$	.1018	^	1	$	&	$	.1854	^	{	4	}	$	&	$	.1812	^	{	4	}	$	&	$	.2681	^	{	4	}	$	&	$	.3234	^	{	4	}	$	&	$	.5870	^	{	4	}	$	&	$	.5440	^	{	4	}	$	\\
$B_5$	&	$	.1019	^	1	$	&	$	.1018	^	1	$	&	$	.1859	^	{	5	}	$	&	$	.1817	^	{	5	}	$	&	$	.2703	^	{	5	}	$	&	$	.3283	^	{	6	}	$	&	$	.6413	^	{	7	}	$	&	$	.5666	^	{	6	}	$	\\
$B_6$	&	$	.1019	^	1	$	&	$	.1018	^	1	$	&	$	.1895	^	{	7	}	$	&	$	.1822	^	{	6	}	$	&	$	.2744	^	{	7	}	$	&	$	.3278	^	{	5	}	$	&	$	.6262	^	{	5	}	$	&	$	.5538	^	{	5	}	$	\\
$B_7$	&	$	.1019	^	1	$	&	$	.1018	^	1	$	&	$	.1864	^	{	6	}	$	&	$	.1822	^	{	6	}	$	&	$	.2711	^	{	6	}	$	&	$	.3300	^	{	7	}	$	&	$	.6396	^	{	6	}	$	&	$	.5680	^	{	7	}	$	\\
$B_8$	&	$	.1019	^	1	$	&	$	.1018	^	1	$	&	$	.1900	^	{	8	}	$	&	$	.1827	^	{	8	}	$	&	$	.2774	^	{	8	}	$	&	$	.3327	^	{	8	}	$	&	$	.6787	^	{	8	}	$	&	$	.5765	^	{	8	}	$	\\
$B_9$	&	$	.1019	^	1	$	&	$	.1018	^	1	$	&	$	.1905	^	{	9	}	$	&	$	.1831	^	{	9	}	$	&	$	.2786	^	{	9	}	$	&	$	.3343	^	{	9	}	$	&	$	.6844	^	{	9	}	$	&	$	.5778	^	{	9	}	$	\\
$B_{10}$	&	$	.1019	^	1	$	&	$	.1018	^	1	$	&	$	.1909	^	{	10	}	$	&	$	.1836	^	{	10	}	$	&	$	.2821	^	{	10	}	$	&	$	.3392	^	{	11	}	$	&	$	.8324	^	{	12	}	$	&	$	.6005	^	{	11	}	$	\\
$B_{11}$	&	$	.1019	^	1	$	&	$	.1018	^	1	$	&	$	.1945	^	{	11	}	$	&	$	.1841	^	{	11	}	$	&	$	.2850	^	{	11	}	$	&	$	.3387	^	{	10	}	$	&	$	.7052	^	{	10	}	$	&	$	.5877	^	{	10	}	$	\\
$B_{12}$	&	$	.1019	^	1	$	&	$	.1018	^	1	$	&	$	.1950	^	{	12	}	$	&	$	.1846	^	{	12	}	$	&	$	.2890	^	{	12	}	$	&	$	.3436	^	{	12	}	$	&	$	.7683	^	{	11	}	$	&	$	.6104	^	{	12	}	$	\\
\hline																																																													
Corr.	&		\multicolumn{2}{c}{1}										&		\multicolumn{2}{c}{0.997}														&		\multicolumn{2}{c}{0.972}														&		\multicolumn{2}{c}{0.986}														\\
\hline																																																													
\end{tabular}																																																																				
\end{center}
\end{table}

\begin{figure}
\center
\includegraphics[width=4in]{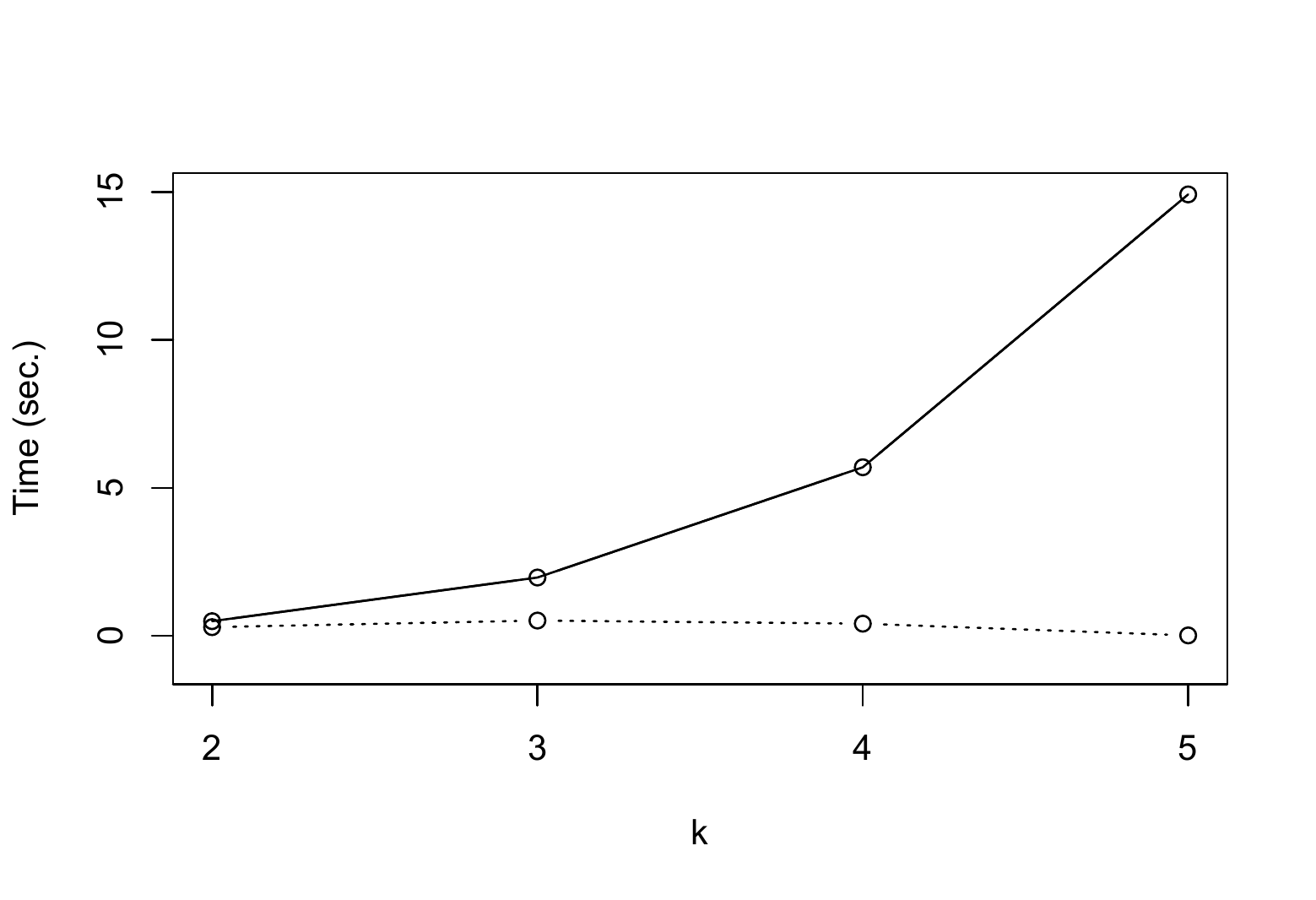}
\caption{Computation time for calculating the average ${P}_{.5}$ values (solid line) and the average ${\tilde P}_{.5}$ values (dotted line) of the designs in Example 4.2.1.}
\label{fi:time}
\end{figure}

{\bf Example 4.2.1.}  
Table~\ref{tb:ex4.2.1} lists twelve nonisomorphic nonregular designs $B_1,\cdots,B_{12}$, which are obtained from the projections of the $14\times 13$ designs in Table~6 in Lin (1993). 
We study the cases that the maximal models of interest are second-order models of $k$ factors for $k=2,\cdots,5$. 
For each design, we calculate the average $P_{.5}$ value and the average ${\tilde P}_{.5}$ value over all $k$-factor projections of the design. 
These average values are shown in Table~\ref{tb:non}. Note that, when $k=4,5$, there are inestimable submodels and the average $P_{.5}'$ values are calculated instead. 
The superscripts in Table~\ref{tb:non} indicate the ranks of the twelve designs determined by the two criteria according to the average ${\tilde P}_{.5}$ values and the average ${P}_{.5}$ (or ${P}_{.5}'$) values. The design with lower rank is better than the design with higher rank. 
Correlations between the ranks given by the two criteria are calculated and listed in the last row. 
Results show that all of the correlations are greater than 0.97 and the top four designs selected by the two criteria are identical. 
Figure~\ref{fi:time} is the comparison of computation time for calculating the average ${P}_{.5}$ values (solid line) and the average ${\tilde P}_{.5}$ values (dotted line) of the twelve designs. 
It shows that computation time for the average ${{ P}}_{.5}$ values increases exponentially when $k$ increases. That is because getting exact average ${{ P}}_{.5}$ values is seriously affected by the total number of eligible submodels ($n_0=5$ as $k=2$, $n_0=18$ as $k=3$, $n_0=113$ as $k=4$, and $n_0=1439$ as $k=5$). On the contrary, calculating the average ${\tilde P}_{.5}$ values is fast and not affected by $n_0$.  
When $k=5$, computation time for the average ${{ P}}_{.5}$ values (14.92 sec.) is nearly 1000 times of that for the average ${\tilde P}_{.5}$ values (0.15 sec.).  
This example shows that the ${{\tilde P}}_{\alpha}$ criterion is computationally inexpensive and has very consistent result with the $P_{\alpha}$ criterion. 
  


\subsection{Saturated and supersaturated designs}
An $N$-run and $m$-factor factorial design is called the saturated design if $m=N-1$ and the supersaturated design if $m\geq N$. 
Since saturated and supersaturated designs can investigate a large number of potential factors with a small run size, they can save considerable cost and are commonly suggested for screening experiments. 
We apply our method to find the ${\tilde P}_\alpha$-optimal saturated designs and give an example to show the connection between the ${\tilde P}_\alpha$ criterion and the $E(s^2)$ criterion for supersaturated designs.

\begin{table}
\begin{center}
\caption{Columnwise-pairwise algorithm for constructing and selecting robust designs}\label{tb:al}
\begin{tabular}{p{15cm}}
\hline
{\bf Algorithm:}\\
\hline
\begin{enumerate}
\item[Step 1.]
\begin{itemize} 
\item[(a)] Start with a level-balanced $N\times m$ design (or near level-balanced for odd $N$), denoted by $D^*$. 
\item[(b)] Calculate ${\tilde P}_\alpha$ for all $k$-factor projections of $D^*$ and take the average, denoted by ${\bar{\tilde P}}_\alpha^*$. 
\end{itemize}
\item[Step 2.] 
\begin{itemize} 
\item[(a)] Reorder columns of $D^*$ by $d_{(1)},\cdots,d_{(m)}$ with $i<j$ iff ${\tilde P}_\alpha(D_{-d(i)}^*)<{\tilde P}_\alpha(D_{-d(j)}^*)$, where $D_{-d(l)}^*$ denotes the $N\times (m-1)$ matrix by deleting column $d_{(l)}$ of $D^*$. 
\item[(b)] Update $D^*$ with the reordered design.
\end{itemize} 
\item[Step 3.] 
\begin{itemize}
\item[(a)] Set $i=1$.
\item[(b)] Conduct a first-order adjustment (see Li and Wu, 1997) for the $i$th column of $D^*$ by changing one element of $d_{(i)}$ from $-1$ to 1 and another element of $d_{(i)}$ from 1 to $-1$ to obtain an updated design $D^{up}$. 
\item[(c)] Calculate ${\tilde P}_\alpha$ for all $k$-factor projections of $D^{up}$ and take the average, denoted by ${\bar{\tilde P}}_\alpha^{up}$. 
\item[(d)] If ${\bar{\tilde P}}_\alpha^{up}<{\bar{\tilde P}}_\alpha^{*}$, replace $D^*$ with $D^{up}$, update ${\bar{\tilde P}}_\alpha^{*}$ with ${\bar{\tilde P}}_\alpha^{up}$, and go to Step 2. If not, go to Step~3~(b) to conduct another first-order adjustment for the $i$th column of $D^*$ until all possible adjustments for column $i$ have been done.
\item[(e)] If $i<g$, where $g$ ($g=5$ is suggested by Li and Wu, 1997) indicates the number of columns of $D^*$ in which the first-order adjustments are conducted, then $i=i+1$ and go to Step~3~(b). Otherwise, stop the procedure. 
\end{itemize}\end{enumerate}\\
\hline
\end{tabular}
\end{center}
\end{table}

\begin{table}
\begin{center}
\caption{Comparison of the $p$-efficient designs and the ${\tilde P}_\alpha$-optimal designs}\label{tb:sa}
	\begin{tabular}{c|ccccc|rrrr}																										
	\hline																										
	Design		&	$k=2$	&	$k=3$	&	$k=4$	&	$k=5$	&	$k=m$	&			\multicolumn{4}{c}{GWLP: $(b_1,b_2,b_3,b_4)$}											\\
	\hline																										
$	L_6	$	&	.2076	&	.2936	&	.3789	&	.4520	&	.4520	&		(	0.00	,	&	1.11	,	&	2.67	,	&	0.56	)	\\
$	D_6	$	&	.2076	&	.2928	&	.3768	&	.4487	&	.4487	&		(	0.00	,	&	1.11	,	&	2.22	,	&	0.56	)	\\
$	L_{10}	$	&	.1217	&	.1666	&	.2198	&	.2809	&	.5093	&		(	0.00	,	&	1.44	,	&	10.08	,	&	14.64	)	\\
$	D_{10}	$	&	.1217	&	.1666	&	.2197	&	.2807	&	.5085	&		(	0.00	,	&	1.44	,	&	9.92	,	&	14.96	)	\\
$	L_{17}	$	&	.0712	&	.0962	&	.1245	&	.1569	&	.6251	&		(	0.06	,	&	1.52	,	&	39.47	,	&	123.11	)	\\
$	D_{17}	$	&	.0711	&	.0958	&	.1238	&	.1557	&	.6146	&		(	0.06	,	&	0.97	,	&	39.36	,	&	124.22	)	\\
$	L_{18}	$	&	.0670	&	.0903	&	.1168	&	.1469	&	.6339	&		(	0.00	,	&	1.68	,	&	43.85	,	&	148.10	)	\\
$	D_{18}	$	&	.0670	&	.0903	&	.1168	&	.1468	&	.6329	&		(	0.00	,	&	1.68	,	&	43.51	,	&	148.00	)	\\
$	L_{21}	$	&	.0575	&	.0774	&	.0998	&	.1251	&	.6748	&		(	0.05	,	&	1.66	,	&	62.16	,	&	258.22	)	\\
$	D_{21}	$	&	.0574	&	.0772	&	.0994	&	.1243	&	.6655	&		(	0.05	,	&	0.99	,	&	62.27	,	&	261.25	)	\\
$	L_{22}	$	&	.0547	&	.0737	&	.0950	&	.1190	&	.6893	&		(	0.00	,	&	2.13	,	&	68.79	,	&	299.51	)	\\
$	D_{22}	$	&	.0547	&	.0736	&	.0948	&	.1186	&	.6824	&		(	0.00	,	&	1.74	,	&	68.07	,	&	300.64	)	\\
$	L_{25}	$	&	.0482	&	.0647	&	.0832	&	.1038	&	.7144	&		(	0.04	,	&	1.43	,	&	90.74	,	&	471.00	)	\\
$	D_{25}	$	&	.0482	&	.0647	&	.0831	&	.1036	&	.7107	&		(	0.04	,	&	1.06	,	&	91.02	,	&	472.96	)	\\
	\hline																										
	\end{tabular}																																
\end{center}
\end{table}

{\bf Example 4.3.1.} 
Based on the same idea with ours that a good design should be able to project onto good projections with nice statistical properties, Lin (1993) proposed the $p$-efficient saturated designs. 
They showed that the $p$-efficient saturated designs are efficient for estimating the parameters of projective first-order models. 
Here, we consider the estimation and prediction efficiency simultaneously and apply our method to select saturated designs that are robust to second-order models.       
To construct and search for robust saturated designs, we modify the columnwise-pairwise algorithm proposed in Li and Wu (1997) and list it in Table~\ref{tb:al}.  
The procedure in Table~\ref{tb:al} is repeated several times with random or specified starting designs. The final design $D^*$ which has the smallest average ${\tilde P}_\alpha$ value is selected as the ${\tilde P}_\alpha$-optimal designs. 
We consider that the maximal model of interest is the second-order model of $k$ factors, where $k=5$, and set $\pi_i=.5$, $\pi_{ij}=.25$ for $i,j=1,\cdots,k$, and $\alpha=0.5$. 
%
The ${\tilde P}_\alpha$ values for $k=2,3,4,5,m$ and the generalized wordlength patterns ($b_1,\cdots,b_4$) are calculated.  
Result shows that, for $N\equiv 0$ (Mod 4) and $N\equiv 3$ (Mod 4), the ${\tilde P}_\alpha$-optimal designs we obtained are equivalent to the $p$-efficient designs  in Lin (1993). They have the same ${\tilde P}_\alpha$ values and generalized wordlength patterns. 
In Table~\ref{tb:sa}, we list the $p$-efficient designs in Lin (1993), denoted by $L_N$, and the ${\tilde P}_\alpha$-optimal designs in Appendix B, denoted by $D_N$, which have different ${\tilde P}_\alpha$ values and generalized wordlength patterns. 
Results show that the ${\tilde P}_\alpha$ values of $D_N$ and $L_N$ are very close when $k=2$. 
It is reasonable because when $k=2$, the ${\tilde P}_\alpha$ criterion minimizes the first two terms (the terms with $b_1$ and $b_2$) in Equation (\ref{eq:2nd}). Tsai et al. (2010) pointed out that the $p$-efficiency criterion first minimizes $b_1$, second minimizes max $|a_{ij}|$, and third minimizes $b_2$. Therefore, the $p$-efficiency criterion is closely related to the ${\tilde P}_\alpha$ criterion when the second-order model of $k=2$ factors are considered. 
The differences of the ${\tilde P}_\alpha$ values between $D_N$ and $L_N$ increase when $k$ becomes large. 
The generalized wordlength patterns also show that the ${\tilde P}_\alpha$-optimal designs in Table~\ref{tb:sa} have less aberration than the $p$-efficient designs.

{\bf Example 4.3.2.}  
The algorithm in Table~\ref{tb:al} can be applied for finding robust supersaturated designs by setting $m\geq N$. In this example, we discuss the connection between the ${\tilde P}_\alpha$ criterion and the $E(s^2)$ criterion (Booth and Cox, 1962), which has been considered as a standard criterion to select supersaturated designs in many papers. 
The $E(s^2)$ criterion selects designs by minimizing $E(s^2)=\sum_{i<j}a_{ij}^2/{m\choose 2}$, which is equal to minimizing $\frac{N^2}{{m\choose 2}}b_2$. 
For a level-balanced design, we have $b_1=0$. 
If the maximal model of interest is the first-order model, the ${\tilde P}_\alpha$ criterion selects a design which minimizes $2(1-\frac{2}{3}\alpha)\xi_2b_2$ (see Section~\ref{se:1st}). 
If the maximal model of interest is the second-order models of $k$ factors, where $k=2$, then  the ${\tilde P}_\alpha$ criterion selects a design which minimizes $[2(1-\frac{2\alpha}{3})\xi_{20}+(1+\frac{\alpha}{9})\xi_{21}+2(1-\frac{8\alpha}{9})(k-2)\xi_{32}]b_2$ (see Section~\ref{se:2nd}). 
Under above two maximal models, both $E(s^2)$ and ${\tilde P}_\alpha$ criteria select designs which have minimum $b_2$ and, hence, the two criteria are equivalent.

\section{Concluding remarks}
\label{se:co}
Many methods for finding robust designs to model uncertainty have been proposed. Most of them are based on the $D$ and $A$ criteria, which aim to select designs with high estimation efficiency. However, the methods for selecting robust designs with high prediction efficiency are rarely developed in the literature. 
In this research, we extend the concept of the mean $A_s$-efficiency in Tsai et al. (2000) to propose the $P_\alpha$ criterion (the $A_s$ and $I_s$ criteria are its special cases). By adjusting the parameter $\alpha$, the $P_\alpha$ criterion can be used for selecting robust designs with high estimation, high prediction, or balanced estimation and prediction efficiency for projective submodels of the maximal model. 
Although the $P_\alpha$ criterion in Section~\ref{se:p} is developed under the framework of two-level designs, it can be easily extended for three-level designs by replacing ${\bf G}_s$ in Equation (\ref{eq:Is}) with the matrix ${\bf B}$ in Lin and Po (2015) and replacing ${\bf X}_s$ with the model matrix of three-level designs. 
We apply the approximation approach in Tsai et al. (2000) and extend the $Q$ and $Q_B$ criteria in Tsai et al. (2000, 2007) to develop the ${\tilde P}_\alpha$ criterion (the ${\tilde A}_s$ and ${\tilde I}_s$ criteria are its special cases). 
It is an approximation version of the $P_\alpha$ criterion and allows us to search designs over a wide range of models. 
Example 4.2.1. shows that the ${{\tilde P}}_{\alpha}$ criterion is computationally inexpensive and has very consistent result with the $P_{\alpha}$ criterion for selecting optimal designs. 
To extend the ${\tilde P}_\alpha$ (or ${\tilde I}_s$) criteria for three-level designs is not easy because it requires the approximation of the off-diagonal elements in the inversion of the information matrix and the approach in Tsai et al. (2000) does not work for this. 
Appropriate approximation methods are currently studied and some extensions of the ${\tilde P}_\alpha$ criterion are underdeveloped. 

\section*{Appendix}

\subsection*{Appendix A}
Following the notations and discussions in Tsai and Gilmour (2010), let $\bf X$ be the $N\times (v+1)$ model matrix of the maximal model and $a_{ij}$, $i,j=0,\cdots,v$, be the $(i,j)$ entry of ${\bf X}'{\bf X}$. 
For $i,j=0$, $i$ and $j$ refer to the intercept; for $i,j=1,\cdots,k$, $i$ and $j$ refer to main effects; for $i,j=k+1,\cdots,v$, $i$ and $j$ refer to two-factor interactions. 
Minimizing Equation (\ref{eq:aP}) is equivalent to minimizing the sum of the following terms: 
(a) $\alpha\sum_{j=1}^k\frac{a_{0j}^2}{N^2}p_{0j}$, (b) $(1-
\frac{2}{3}\alpha)\sum_{i=1}^k\frac{a_{i0}^2}{N^2}p_{i0}$, (c) $\alpha\sum_{j=k+1}^v\frac{a_{0j}^2}{N^2}p_{0j}$, (d) $(1-
\frac{8}{9}\alpha)\sum_{i=k+1}^v\frac{a_{i0}^2}{n^2}p_{i0}$, (e) $(1-\frac{2}{3}\alpha)\sum_{i=1}^k\sum_{\substack{j=1\\i\neq j}}^k\frac{a_{ij}^2}{N^2}p_{ij}$, (f) $(1-\frac{8}{9}\alpha)\sum_{i=k+1}^v\sum_{\substack{j=k+1\\i\neq j}}^v\frac{a_{ij}^2}{N^2}p_{ij}$, (g) $(1-\frac{2}{3}\alpha)\sum_{i=1}^k\sum_{j=k+1}^v\frac{a_{ij}^2}{N^2}p_{ij}$, and (h) $(1-\frac{8}{9}\alpha)\sum_{i=k+1}^v\sum_{j=1}^k\frac{a_{ij}^2}{N^2}p_{ij}$. 
Let $\xi_{ab}$ denote the sum of prior probabilities of the models, which include $a$ main effects and $b$ two-factor interactions, being the final model.  
\begin{enumerate}
\item In (a), the sum of $\frac{a_{0j}^2}{N^2}$ is equal to $b_1$ and the $p_{0j}$ associated with the $a_{0j}$ is $\xi_{10}$. Hence, (a) equals $\alpha\xi_{10}b_1$. Similarly, in (b), the sum of $\frac{a_{i0}^2}{N^2}$ is equal to $b_1$ and the $p_{i0}$ associated with the $a_{i0}$ is $\xi_{10}$. Hence, (b) equals $(1-
\frac{2}{3}\alpha)\xi_{10}b_1$. 

\item In (c), the sum of $\frac{a_{0j}^2}{N^2}$ is equal to $b_2$ and the $p_{0j}$ associated with the $a_{0j}$ is $\xi_{21}$. Hence, (c) equals to $\alpha\xi_{21}b_2$. Similarly, in (d), the sum of $\frac{a_{i0}^2}{N^2}$ is equal to $b_2$ and the $p_{i0}$ associated with the $a_{i0}$ is $\xi_{21}$. Hence, (d) equals $(1-
\frac{8}{9}\alpha)\xi_{21}b_2$. 

\item In (e), the sum of $\frac{a_{ij}^2}{N^2}$ is equal to $2b_2$ and the $p_{ij}$ associated with the $a_{ij}$ is $\xi_{20}$. Hence, (e) equals $2(1-\frac{2}{3}\alpha)\xi_{20}b_2$.

\item In (f), for $i,j$ referring to interactions with a common factor, the sum of $\frac{a_{ij}^2}{N^2}$ is equal to $2(k-2)b_2$ and the $p_{ij}$ associated with the $a_{ij}$ is $\xi_{32}$; for $i,j$ referring to interactions with no common factor, the sum of $\frac{a_{ij}^2}{N^2}$ is equal to $6b_4$ and the $p_{ij}$ associated with the $a_{ij}$ is $\xi_{42}$. Hence, (f) equals $2(1-\frac{8}{9}\alpha)(k-2)\xi_{32}b_2+6(1-\frac{8}{9}\alpha)\xi_{42}b_4$.

\item In (g), for $i$ referring to the main effect of a factor and $j$ referring to an interaction, where $i$ and $j$ have a common factor, the sum of $\frac{a_{ij}^2}{N^2}$ is equal to $2(k-2)b_1$ and the $p_{ij}$ associated with the $a_{ij}$ is $\xi_{21}$; for $i$ referring to the main effect of a factor and $j$ referring to an interaction, where $i$ and $j$ have no common factor, the sum of $\frac{a_{ij}^2}{N^2}$ is equal to $6b_3$ and the $p_{ij}$ associated with the $a_{ij}$ is $\xi_{31}$. Hence, (g) equals $2(1-\frac{2}{3}\alpha)(k-2)\xi_{21}b_1+6(1-\frac{2}{3}\alpha)\xi_{31}b_3$. With the same arguments, we obtain that (h) equals $2(1-\frac{8}{9}\alpha)(k-2)\xi_{21}b_1+6(1-\frac{8}{9}\alpha)\xi_{31}b_3$. 
\end{enumerate}
Summarizing above discussions, we obtain Equation (\ref{eq:2nd}).

\subsection*{Appendix B}\label{ap:sa}

\begin{center}
\begin{spacing}{.7}	
\begin{tabular}{c}																																															
$N_6$\\																																															
\begin{tabular}{rrrrr}																																															
\hline																																															
1	&	2	&	3	&	4	&	5	\\																																						
\hline																																															
-1	&	-1	&	1	&	1	&	1	\\																																						
1	&	-1	&	-1	&	1	&	1	\\																																						
1	&	1	&	-1	&	-1	&	1	\\																																						
1	&	1	&	1	&	-1	&	-1	\\																																						
-1	&	1	&	1	&	1	&	-1	\\																																						
-1	&	-1	&	-1	&	-1	&	-1	\\																																						
\hline																																															
\end{tabular}\\																																															
\end{tabular}																																															
\[\]																																															
\begin{tabular}{c}																																															
$N_{10}$\\																																															
\begin{tabular}{rrrrrrrrr}																																															
\hline																																															
1	&	2	&	3	&	4	&	5	&	6	&	7	&	8	&	9	\\																														
\hline																																															
1	&	-1	&	-1	&	-1	&	-1	&	1	&	-1	&	1	&	1	\\																														
1	&	1	&	-1	&	1	&	1	&	-1	&	1	&	-1	&	1	\\																														
1	&	1	&	1	&	1	&	-1	&	1	&	-1	&	-1	&	-1	\\																														
1	&	-1	&	-1	&	1	&	1	&	-1	&	-1	&	1	&	-1	\\																														
1	&	-1	&	1	&	-1	&	1	&	1	&	1	&	1	&	-1	\\																														
-1	&	1	&	1	&	-1	&	1	&	-1	&	-1	&	-1	&	-1	\\																														
-1	&	1	&	-1	&	-1	&	-1	&	-1	&	1	&	1	&	-1	\\																														
-1	&	1	&	1	&	1	&	-1	&	1	&	1	&	1	&	1	\\																														
-1	&	-1	&	1	&	-1	&	-1	&	-1	&	1	&	-1	&	1	\\																														
-1	&	-1	&	-1	&	1	&	1	&	1	&	-1	&	-1	&	1	\\																														
\hline																																															
\end{tabular}\\																																															
\end{tabular}																																															
\[\]																																															
\begin{tabular}{c}																																															
$N_{17}$\\																																															
\begin{tabular}{rrrrrrrrrrrrrrrr}																																															
\hline																																															
1	&	2	&	3	&	4	&	5	&	6	&	7	&	8	&	9	&	10	&	11	&	12	&	13	&	14	&	15	&	16	\\																
\hline																																															
1	&	1	&	1	&	-1	&	1	&	-1	&	1	&	-1	&	1	&	1	&	-1	&	-1	&	1	&	1	&	-1	&	-1	\\																
1	&	-1	&	1	&	1	&	-1	&	-1	&	1	&	-1	&	-1	&	1	&	1	&	1	&	-1	&	1	&	1	&	-1	\\																
-1	&	1	&	-1	&	-1	&	-1	&	1	&	1	&	1	&	-1	&	1	&	-1	&	1	&	1	&	-1	&	1	&	-1	\\																
-1	&	-1	&	-1	&	1	&	-1	&	1	&	-1	&	-1	&	1	&	1	&	-1	&	-1	&	1	&	1	&	1	&	1	\\																
-1	&	1	&	-1	&	1	&	1	&	-1	&	-1	&	-1	&	1	&	-1	&	1	&	1	&	1	&	-1	&	1	&	-1	\\																
-1	&	1	&	1	&	1	&	-1	&	1	&	-1	&	1	&	-1	&	-1	&	1	&	-1	&	1	&	1	&	-1	&	-1	\\																
-1	&	1	&	-1	&	1	&	-1	&	-1	&	1	&	1	&	1	&	-1	&	-1	&	1	&	-1	&	1	&	-1	&	1	\\																
-1	&	1	&	-1	&	-1	&	1	&	1	&	-1	&	-1	&	-1	&	1	&	1	&	1	&	-1	&	1	&	-1	&	1	\\																
1	&	-1	&	1	&	-1	&	1	&	1	&	-1	&	1	&	1	&	-1	&	-1	&	1	&	-1	&	1	&	1	&	-1	\\																
1	&	-1	&	1	&	-1	&	-1	&	1	&	1	&	-1	&	1	&	-1	&	1	&	1	&	1	&	-1	&	-1	&	1	\\																
1	&	-1	&	-1	&	1	&	1	&	1	&	1	&	1	&	1	&	1	&	1	&	-1	&	-1	&	-1	&	-1	&	-1	\\																
1	&	1	&	1	&	-1	&	-1	&	-1	&	-1	&	1	&	1	&	1	&	1	&	-1	&	-1	&	-1	&	1	&	1	\\																
1	&	-1	&	-1	&	-1	&	1	&	-1	&	1	&	1	&	-1	&	-1	&	1	&	-1	&	1	&	1	&	1	&	1	\\																
1	&	-1	&	1	&	1	&	1	&	-1	&	-1	&	1	&	-1	&	1	&	-1	&	1	&	1	&	-1	&	-1	&	1	\\																
-1	&	1	&	1	&	1	&	1	&	1	&	1	&	-1	&	-1	&	-1	&	-1	&	-1	&	-1	&	-1	&	1	&	1	\\																
-1	&	-1	&	-1	&	-1	&	-1	&	-1	&	-1	&	-1	&	-1	&	-1	&	-1	&	-1	&	-1	&	-1	&	-1	&	-1	\\																
1	&	1	&	-1	&	1	&	-1	&	1	&	-1	&	-1	&	-1	&	-1	&	-1	&	-1	&	1	&	1	&	1	&	-1	\\																
\hline																																															
\end{tabular}\\																																															
\end{tabular}																																															
\[\]																																															
\begin{tabular}{c}																																															
$N_{18}$\\																																															
\begin{tabular}{rrrrrrrrrrrrrrrrr}																																															
\hline																																															
1	&	2	&	3	&	4	&	5	&	6	&	7	&	8	&	9	&	10	&	11	&	12	&	13	&	14	&	15	&	16	&	17	\\														
\hline																																															
1	&	1	&	1	&	1	&	1	&	1	&	1	&	1	&	1	&	1	&	-1	&	1	&	1	&	1	&	1	&	1	&	1	\\														
1	&	1	&	-1	&	1	&	-1	&	-1	&	-1	&	-1	&	-1	&	1	&	1	&	1	&	1	&	1	&	1	&	1	&	1	\\														
1	&	-1	&	-1	&	1	&	1	&	1	&	-1	&	-1	&	1	&	-1	&	1	&	1	&	1	&	-1	&	1	&	-1	&	-1	\\														
-1	&	1	&	-1	&	1	&	-1	&	-1	&	1	&	1	&	1	&	-1	&	-1	&	-1	&	1	&	-1	&	-1	&	1	&	1	\\														
-1	&	1	&	1	&	-1	&	1	&	1	&	-1	&	-1	&	-1	&	1	&	-1	&	1	&	1	&	-1	&	-1	&	-1	&	1	\\														
-1	&	-1	&	1	&	1	&	-1	&	1	&	-1	&	-1	&	1	&	1	&	-1	&	-1	&	1	&	1	&	-1	&	1	&	-1	\\														
1	&	-1	&	-1	&	-1	&	1	&	1	&	1	&	1	&	-1	&	1	&	1	&	-1	&	1	&	-1	&	-1	&	1	&	-1	\\														
-1	&	-1	&	1	&	-1	&	-1	&	-1	&	1	&	1	&	1	&	-1	&	1	&	1	&	1	&	1	&	1	&	-1	&	1	\\														
1	&	1	&	1	&	-1	&	1	&	-1	&	-1	&	1	&	-1	&	-1	&	-1	&	-1	&	1	&	1	&	1	&	-1	&	-1	\\														
-1	&	1	&	1	&	-1	&	-1	&	1	&	1	&	-1	&	-1	&	-1	&	1	&	-1	&	-1	&	-1	&	1	&	1	&	-1	\\														
-1	&	1	&	-1	&	-1	&	1	&	-1	&	1	&	-1	&	1	&	1	&	1	&	-1	&	-1	&	1	&	-1	&	-1	&	-1	\\														
1	&	-1	&	1	&	-1	&	-1	&	-1	&	1	&	-1	&	1	&	-1	&	-1	&	1	&	-1	&	-1	&	-1	&	-1	&	-1	\\														
1	&	-1	&	-1	&	1	&	1	&	1	&	1	&	-1	&	-1	&	-1	&	-1	&	-1	&	-1	&	1	&	1	&	-1	&	1	\\														
-1	&	1	&	-1	&	1	&	-1	&	1	&	1	&	1	&	-1	&	1	&	-1	&	1	&	-1	&	-1	&	1	&	-1	&	-1	\\														
-1	&	-1	&	1	&	1	&	1	&	-1	&	-1	&	1	&	-1	&	-1	&	1	&	1	&	-1	&	1	&	-1	&	1	&	-1	\\														
1	&	1	&	-1	&	-1	&	-1	&	1	&	-1	&	1	&	1	&	-1	&	1	&	1	&	-1	&	1	&	-1	&	1	&	1	\\														
-1	&	-1	&	-1	&	-1	&	1	&	-1	&	-1	&	-1	&	1	&	1	&	-1	&	-1	&	-1	&	-1	&	1	&	1	&	1	\\														
1	&	-1	&	1	&	1	&	-1	&	-1	&	-1	&	1	&	-1	&	1	&	1	&	-1	&	-1	&	-1	&	-1	&	-1	&	1	\\														
\hline																																															
\end{tabular}\\																																															
\end{tabular}																																															
\[\]																																															
\begin{tabular}{c}																																															
$N_{21}$\\																																															
\begin{tabular}{rrrrrrrrrrrrrrrrrrrr}																																															
\hline																																															
1	&	2	&	3	&	4	&	5	&	6	&	7	&	8	&	9	&	10	&	11	&	12	&	13	&	14	&	15	&	16	&	17	&	18	&	19	&	20	\\								
\hline																																															
-1	&	1	&	1	&	-1	&	-1	&	-1	&	1	&	-1	&	-1	&	1	&	1	&	1	&	-1	&	1	&	-1	&	1	&	1	&	1	&	-1	&	-1	\\								
-1	&	1	&	1	&	1	&	-1	&	1	&	1	&	1	&	1	&	-1	&	-1	&	1	&	1	&	-1	&	-1	&	-1	&	1	&	-1	&	-1	&	-1	\\								
-1	&	-1	&	1	&	1	&	-1	&	1	&	1	&	-1	&	-1	&	1	&	-1	&	-1	&	1	&	-1	&	-1	&	1	&	-1	&	1	&	1	&	1	\\								
-1	&	-1	&	-1	&	1	&	-1	&	-1	&	1	&	1	&	1	&	-1	&	1	&	-1	&	-1	&	-1	&	1	&	1	&	1	&	1	&	-1	&	1	\\								
-1	&	1	&	-1	&	1	&	1	&	-1	&	-1	&	1	&	-1	&	1	&	1	&	-1	&	1	&	-1	&	-1	&	1	&	1	&	-1	&	1	&	-1	\\								
-1	&	1	&	1	&	-1	&	-1	&	-1	&	-1	&	1	&	1	&	1	&	-1	&	-1	&	1	&	1	&	1	&	1	&	-1	&	-1	&	-1	&	1	\\								
-1	&	-1	&	1	&	-1	&	1	&	-1	&	1	&	1	&	1	&	1	&	1	&	1	&	-1	&	-1	&	-1	&	-1	&	-1	&	-1	&	1	&	1	\\								
-1	&	1	&	-1	&	1	&	-1	&	1	&	1	&	-1	&	1	&	1	&	1	&	-1	&	-1	&	1	&	1	&	-1	&	-1	&	-1	&	1	&	-1	\\								
-1	&	1	&	1	&	1	&	1	&	-1	&	-1	&	-1	&	1	&	-1	&	-1	&	1	&	-1	&	-1	&	1	&	1	&	-1	&	1	&	1	&	-1	\\								
1	&	-1	&	1	&	-1	&	1	&	1	&	1	&	1	&	-1	&	-1	&	-1	&	-1	&	-1	&	1	&	1	&	1	&	1	&	-1	&	1	&	-1	\\								
1	&	-1	&	-1	&	1	&	1	&	-1	&	1	&	1	&	-1	&	1	&	-1	&	1	&	1	&	1	&	1	&	-1	&	-1	&	1	&	-1	&	-1	\\								
1	&	-1	&	-1	&	1	&	1	&	1	&	-1	&	-1	&	1	&	1	&	-1	&	1	&	-1	&	1	&	-1	&	1	&	1	&	-1	&	-1	&	1	\\								
1	&	-1	&	-1	&	-1	&	-1	&	1	&	-1	&	1	&	1	&	-1	&	1	&	1	&	1	&	1	&	-1	&	1	&	-1	&	1	&	1	&	-1	\\								
1	&	1	&	-1	&	-1	&	-1	&	1	&	-1	&	1	&	-1	&	1	&	-1	&	1	&	-1	&	-1	&	1	&	-1	&	1	&	1	&	1	&	1	\\								
1	&	-1	&	1	&	-1	&	1	&	1	&	-1	&	-1	&	1	&	1	&	1	&	-1	&	1	&	-1	&	1	&	-1	&	1	&	1	&	-1	&	-1	\\								
1	&	1	&	-1	&	-1	&	1	&	-1	&	1	&	-1	&	1	&	-1	&	-1	&	-1	&	1	&	1	&	-1	&	-1	&	1	&	1	&	1	&	1	\\								
1	&	-1	&	1	&	1	&	-1	&	-1	&	-1	&	-1	&	-1	&	-1	&	1	&	1	&	1	&	1	&	1	&	-1	&	1	&	-1	&	1	&	1	\\								
1	&	1	&	-1	&	-1	&	1	&	1	&	1	&	-1	&	-1	&	-1	&	1	&	1	&	1	&	-1	&	1	&	1	&	-1	&	-1	&	-1	&	1	\\								
-1	&	1	&	1	&	1	&	1	&	1	&	-1	&	1	&	-1	&	-1	&	1	&	-1	&	-1	&	1	&	-1	&	-1	&	-1	&	1	&	-1	&	1	\\								
1	&	-1	&	-1	&	-1	&	-1	&	-1	&	-1	&	-1	&	-1	&	-1	&	-1	&	-1	&	-1	&	-1	&	-1	&	-1	&	-1	&	-1	&	-1	&	-1	\\								
1	&	1	&	1	&	1	&	-1	&	-1	&	1	&	1	&	1	&	1	&	1	&	-1	&	-1	&	-1	&	-1	&	1	&	-1	&	-1	&	-1	&	-1	\\								
\hline																																															
\end{tabular}\\																																															
\end{tabular}																																															
\[\]																																															
\begin{tabular}{c}																																															
$N_{22}$\\																																															
\begin{tabular}{rrrrrrrrrrrrrrrrrrrrr}																																															
\hline																																															
1	&	2	&	3	&	4	&	5	&	6	&	7	&	8	&	9	&	10	&	11	&	12	&	13	&	14	&	15	&	16	&	17	&	18	&	19	&	20	&	21	\\						
\hline																																															
-1	&	1	&	1	&	1	&	1	&	1	&	1	&	-1	&	1	&	1	&	1	&	1	&	-1	&	1	&	1	&	1	&	1	&	1	&	1	&	1	&	-1	\\						
1	&	1	&	1	&	1	&	1	&	1	&	-1	&	1	&	1	&	1	&	1	&	1	&	1	&	-1	&	1	&	1	&	1	&	1	&	-1	&	-1	&	1	\\						
1	&	1	&	1	&	-1	&	1	&	1	&	1	&	1	&	-1	&	1	&	-1	&	-1	&	1	&	-1	&	1	&	-1	&	1	&	-1	&	1	&	1	&	1	\\						
-1	&	-1	&	1	&	-1	&	1	&	-1	&	1	&	1	&	-1	&	-1	&	1	&	1	&	1	&	1	&	-1	&	1	&	-1	&	-1	&	1	&	-1	&	-1	\\						
1	&	1	&	-1	&	-1	&	1	&	-1	&	1	&	-1	&	1	&	-1	&	-1	&	-1	&	1	&	1	&	1	&	1	&	1	&	-1	&	-1	&	-1	&	-1	\\						
1	&	-1	&	-1	&	1	&	1	&	1	&	-1	&	-1	&	-1	&	1	&	1	&	-1	&	1	&	-1	&	-1	&	1	&	1	&	-1	&	1	&	1	&	-1	\\						
-1	&	-1	&	-1	&	1	&	1	&	-1	&	-1	&	1	&	1	&	-1	&	-1	&	1	&	-1	&	-1	&	1	&	-1	&	1	&	1	&	1	&	-1	&	-1	\\						
-1	&	1	&	-1	&	-1	&	1	&	1	&	-1	&	-1	&	-1	&	1	&	-1	&	1	&	1	&	1	&	-1	&	-1	&	-1	&	1	&	-1	&	1	&	-1	\\						
-1	&	-1	&	-1	&	-1	&	1	&	1	&	1	&	1	&	1	&	-1	&	1	&	-1	&	-1	&	-1	&	-1	&	1	&	-1	&	1	&	-1	&	1	&	1	\\						
-1	&	1	&	-1	&	1	&	1	&	-1	&	-1	&	-1	&	1	&	1	&	1	&	-1	&	-1	&	1	&	-1	&	-1	&	-1	&	-1	&	1	&	-1	&	1	\\						
1	&	-1	&	1	&	1	&	1	&	-1	&	-1	&	-1	&	-1	&	-1	&	-1	&	-1	&	-1	&	1	&	1	&	-1	&	-1	&	1	&	-1	&	1	&	1	\\						
1	&	1	&	-1	&	1	&	-1	&	1	&	1	&	1	&	-1	&	-1	&	1	&	-1	&	-1	&	1	&	-1	&	-1	&	1	&	1	&	-1	&	-1	&	-1	\\						
1	&	-1	&	1	&	-1	&	-1	&	-1	&	-1	&	1	&	1	&	1	&	-1	&	1	&	-1	&	1	&	-1	&	1	&	1	&	-1	&	-1	&	1	&	1	\\						
1	&	1	&	1	&	-1	&	-1	&	-1	&	-1	&	1	&	-1	&	1	&	-1	&	-1	&	-1	&	-1	&	-1	&	1	&	-1	&	1	&	1	&	-1	&	-1	\\						
1	&	-1	&	1	&	1	&	-1	&	1	&	1	&	-1	&	1	&	-1	&	-1	&	1	&	-1	&	-1	&	-1	&	-1	&	-1	&	-1	&	1	&	1	&	-1	\\						
1	&	-1	&	-1	&	-1	&	-1	&	-1	&	1	&	1	&	1	&	1	&	1	&	1	&	1	&	1	&	1	&	-1	&	-1	&	1	&	1	&	1	&	1	\\						
1	&	1	&	-1	&	-1	&	-1	&	1	&	-1	&	-1	&	-1	&	-1	&	1	&	1	&	-1	&	-1	&	1	&	1	&	-1	&	-1	&	1	&	-1	&	1	\\						
-1	&	-1	&	-1	&	1	&	-1	&	1	&	1	&	1	&	-1	&	1	&	-1	&	-1	&	1	&	1	&	1	&	1	&	-1	&	-1	&	-1	&	-1	&	1	\\						
-1	&	1	&	1	&	1	&	-1	&	-1	&	-1	&	1	&	1	&	-1	&	1	&	-1	&	1	&	-1	&	1	&	-1	&	-1	&	-1	&	-1	&	1	&	-1	\\						
-1	&	-1	&	1	&	-1	&	-1	&	-1	&	1	&	-1	&	-1	&	1	&	1	&	1	&	-1	&	-1	&	1	&	-1	&	1	&	-1	&	-1	&	-1	&	-1	\\						
-1	&	1	&	-1	&	1	&	-1	&	-1	&	1	&	-1	&	-1	&	-1	&	-1	&	1	&	1	&	-1	&	-1	&	1	&	1	&	1	&	-1	&	1	&	1	\\						
-1	&	-1	&	1	&	-1	&	-1	&	1	&	-1	&	-1	&	1	&	-1	&	-1	&	-1	&	1	&	1	&	-1	&	-1	&	1	&	1	&	1	&	-1	&	1	\\						
\hline																																															
\end{tabular}\\																																															
\end{tabular}																																															
\[\]																																															
\begin{tabular}{c}																																															
$N_{25}$\\	
{\scriptsize																																														
\begin{tabular}{rrrrrrrrrrrrrrrrrrrrrrrr}																																															
\hline																																															
1	&	2	&	3	&	4	&	5	&	6	&	7	&	8	&	9	&	10	&	11	&	12	&	13	&	14	&	15	&	16	&	17	&	18	&	19	&	20	&	21	&	22	&	23	&	24	\\
\hline																																															
-1	&	-1	&	-1	&	1	&	-1	&	1	&	1	&	-1	&	1	&	1	&	1	&	1	&	-1	&	1	&	-1	&	-1	&	-1	&	1	&	-1	&	1	&	-1	&	1	&	-1	&	1	\\
-1	&	-1	&	1	&	1	&	1	&	1	&	1	&	1	&	1	&	-1	&	-1	&	-1	&	1	&	1	&	1	&	-1	&	-1	&	1	&	-1	&	-1	&	1	&	-1	&	-1	&	-1	\\
-1	&	1	&	1	&	1	&	-1	&	1	&	1	&	1	&	-1	&	-1	&	1	&	-1	&	-1	&	-1	&	-1	&	-1	&	1	&	-1	&	-1	&	1	&	1	&	-1	&	1	&	1	\\
1	&	1	&	-1	&	1	&	-1	&	1	&	-1	&	-1	&	1	&	-1	&	1	&	1	&	1	&	-1	&	1	&	1	&	1	&	-1	&	-1	&	-1	&	1	&	1	&	-1	&	-1	\\
-1	&	-1	&	1	&	-1	&	1	&	1	&	-1	&	-1	&	-1	&	-1	&	1	&	1	&	-1	&	1	&	1	&	-1	&	-1	&	-1	&	1	&	1	&	1	&	1	&	1	&	-1	\\
1	&	1	&	-1	&	-1	&	1	&	-1	&	-1	&	1	&	1	&	1	&	1	&	-1	&	-1	&	1	&	1	&	1	&	-1	&	-1	&	-1	&	1	&	1	&	-1	&	-1	&	1	\\
-1	&	-1	&	1	&	-1	&	-1	&	-1	&	-1	&	1	&	1	&	-1	&	1	&	-1	&	1	&	-1	&	1	&	-1	&	1	&	1	&	1	&	1	&	-1	&	1	&	-1	&	1	\\
-1	&	1	&	1	&	-1	&	-1	&	-1	&	1	&	-1	&	1	&	1	&	-1	&	1	&	1	&	1	&	1	&	-1	&	1	&	-1	&	-1	&	1	&	-1	&	-1	&	1	&	-1	\\
1	&	1	&	1	&	1	&	1	&	-1	&	-1	&	1	&	1	&	-1	&	-1	&	1	&	-1	&	-1	&	-1	&	1	&	-1	&	1	&	-1	&	1	&	-1	&	1	&	1	&	-1	\\
-1	&	1	&	1	&	-1	&	1	&	1	&	1	&	-1	&	1	&	-1	&	-1	&	-1	&	-1	&	1	&	-1	&	1	&	1	&	-1	&	1	&	-1	&	-1	&	1	&	-1	&	1	\\
1	&	1	&	1	&	1	&	-1	&	-1	&	-1	&	1	&	-1	&	1	&	-1	&	1	&	1	&	1	&	-1	&	-1	&	-1	&	-1	&	1	&	-1	&	1	&	1	&	-1	&	1	\\
1	&	1	&	-1	&	-1	&	1	&	1	&	-1	&	1	&	-1	&	1	&	1	&	-1	&	1	&	1	&	-1	&	-1	&	1	&	1	&	-1	&	-1	&	-1	&	1	&	1	&	-1	\\
-1	&	-1	&	-1	&	-1	&	-1	&	-1	&	1	&	1	&	-1	&	-1	&	-1	&	1	&	-1	&	1	&	1	&	1	&	1	&	1	&	-1	&	-1	&	1	&	1	&	1	&	1	\\
1	&	-1	&	-1	&	1	&	1	&	-1	&	1	&	1	&	-1	&	-1	&	1	&	1	&	1	&	1	&	-1	&	1	&	1	&	-1	&	1	&	1	&	-1	&	-1	&	-1	&	-1	\\
1	&	-1	&	-1	&	1	&	1	&	1	&	-1	&	1	&	1	&	1	&	-1	&	1	&	-1	&	-1	&	1	&	-1	&	1	&	-1	&	1	&	-1	&	-1	&	-1	&	1	&	1	\\
-1	&	-1	&	1	&	-1	&	1	&	1	&	-1	&	-1	&	-1	&	1	&	-1	&	1	&	1	&	-1	&	-1	&	1	&	1	&	1	&	-1	&	1	&	1	&	-1	&	-1	&	1	\\
1	&	1	&	-1	&	-1	&	1	&	-1	&	1	&	-1	&	1	&	-1	&	1	&	1	&	1	&	-1	&	-1	&	-1	&	-1	&	1	&	1	&	-1	&	1	&	-1	&	1	&	1	\\
1	&	-1	&	1	&	1	&	1	&	-1	&	1	&	-1	&	-1	&	1	&	1	&	-1	&	1	&	-1	&	1	&	1	&	-1	&	-1	&	-1	&	-1	&	-1	&	1	&	1	&	1	\\
1	&	1	&	-1	&	1	&	-1	&	1	&	-1	&	-1	&	-1	&	-1	&	-1	&	-1	&	1	&	1	&	1	&	1	&	-1	&	1	&	1	&	1	&	-1	&	-1	&	1	&	1	\\
-1	&	-1	&	-1	&	-1	&	-1	&	1	&	1	&	1	&	1	&	1	&	-1	&	-1	&	1	&	-1	&	-1	&	1	&	-1	&	-1	&	1	&	1	&	1	&	1	&	1	&	-1	\\
-1	&	-1	&	1	&	1	&	-1	&	-1	&	-1	&	-1	&	1	&	1	&	1	&	-1	&	-1	&	1	&	-1	&	1	&	1	&	1	&	1	&	-1	&	1	&	-1	&	1	&	-1	\\
1	&	1	&	1	&	-1	&	-1	&	1	&	1	&	1	&	-1	&	1	&	1	&	1	&	-1	&	-1	&	1	&	1	&	-1	&	1	&	1	&	-1	&	-1	&	-1	&	-1	&	-1	\\
1	&	1	&	-1	&	1	&	1	&	-1	&	1	&	-1	&	-1	&	1	&	-1	&	-1	&	-1	&	-1	&	1	&	-1	&	1	&	1	&	1	&	1	&	1	&	1	&	-1	&	-1	\\
-1	&	-1	&	-1	&	-1	&	-1	&	-1	&	-1	&	-1	&	-1	&	-1	&	-1	&	-1	&	-1	&	-1	&	-1	&	-1	&	-1	&	-1	&	-1	&	-1	&	-1	&	-1	&	-1	&	-1	\\
1	&	-1	&	1	&	-1	&	-1	&	1	&	1	&	-1	&	1	&	-1	&	-1	&	-1	&	-1	&	1	&	-1	&	-1	&	1	&	-1	&	-1	&	1	&	1	&	1	&	-1	&	-1	\\
\hline																																															
\end{tabular}}\\																																															
\end{tabular}\\																																																																																																																																									
\end{spacing}										
\end{center}

\end{document}